\documentclass[10pt]{article}
\usepackage{graphics, epsfig,colordvi,afterpage}
\usepackage[round, authoryear]{natbib}
\usepackage{geometry}
\usepackage{import}

\usepackage{amssymb}
\usepackage{amsmath}
\usepackage{textcomp}
\usepackage{gensymb}
\usepackage{cancel}
\usepackage{lineno}
\usepackage{xcolor}
\usepackage{hyperref}
\hypersetup{colorlinks=true, linkcolor=red, citecolor = cyan}
\usepackage{soul}
\usepackage{verbatim}
\usepackage{placeins}
\usepackage{ulem}
\usepackage{array}
\usepackage{tabularx}
\usepackage{multirow}

\begin{document}

\fontsize{10pt}{18pt}\selectfont

\centerline{The dominance of feedback mechanisms for perception in the central visual field:}
\centerline{the flip tilt illusion by backward masking and the Feedforward-Feedback-Verify-reWeight algorithm}

\vspace { 2 cm}

\centerline{Li Zhaoping}

\centerline{University of T\"ubingen and Max Planck Institute for Biological Cybernetics}
\centerline{email: li.zhaoping@tuebingen.mpg.de}

\section*{Abstract}
The flip tilt illusion is the misperception of the orientation of the alignment between a pair
of dots by 90 degrees. It occurs when the dot-pair contains one black dot and one white dot (in a gray background), 
is typically visible only in the peripheral visual field, and is often demonstrated when many pairs of dots
are grouped into a texture or a contour.  
This illusion is predicted by the central-peripheral dichotomy (CPD) theory as follows.
Along the visual pathway from retina to downstream stages, our brain admits only a tiny fraction of visual input 
information for recognition due to a processing bottleneck that starts from the output of the primary visual cortex (V1) along the visual pathway. 
Accordingly,  perception downstream from V1 relies on impoverished information,
selected largely by gaze shifts to center the selected content at fovea and additionally
by the capacity of the bottleneck even in the central visual field.
Based on impoverished information initially fed forward 
along the visual pathway, initial perceptual hypotheses about the visual scene 
can be ambiguous.  Illusions can occur when the impoverished information is additionally misleading.
To overcome, downstream areas send feedback to upstream areas such as 
V1 to query for additional information, thereby verifying and re-assessing the weights of the initial hypotheses.  
This perceptual algorithm is termed Feedforward-Feedback-Verify-reWeight (FFVW).
To save resources, the feedback query for Feedback-Verify-reWeight (FVW) is limited largely 
to the central visual field.  The peripheral visual field is thus vulnerable to not only 
perceptual ambiguities, but also to illusions, as demonstrated in the flip tilt illusion. 
The CPD theory further predicts that impairing the feedback renders the illusion visible in the central visual field.  
This paper reports a confirmation of the prediction in an experiment that disrupted the feedback by masking 
the visual inputs after a short viewing duration $t$.
Furthermore, the FFVW algorithm is studied by a computational model of the perceptual data as a function 
of $t$.  It reveals that the initial feedforward sweep of visual information 
explains the flip tilt illusion for $t \le 20$ ms, whereas the feedback mechanism dominates perception 
for longer $t$'s.  Future studies should apply the model to 
other visual perceptual tasks, thereby quantitatively and phenomenologically assess
the task-dependent, feedforward and feedback, mechanisms.

\section*{Keywords}
Central-Peripheral Dichotomy (CPD);  the CPD theory ;  Central vision  ;    Peripheral vision ;    
feedforward ;    feedback ;    backward masking ;    recognition ;    analysis-by-synthesis  ;    hypothesis testing; the Feedforward-Feedback-Verify-reWeight (FFVW) algorithm

\section{Introduction} \label{sec:introduction}

\subsection{The flip tilt illusion, the bottleneck,  and the Central-Peripheral Dichotomy (CPD) theory}

\begin{figure}[hhhh!]
\centering
\includegraphics[width=\textwidth]{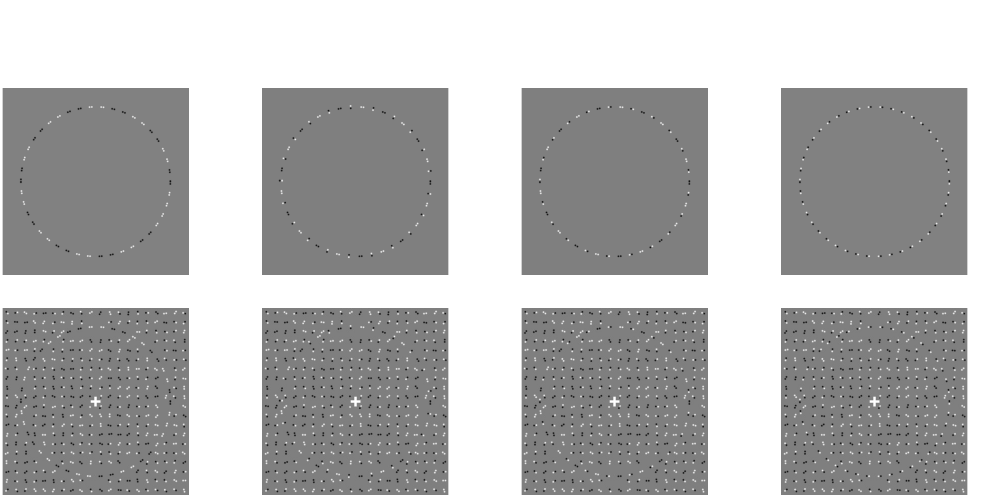}
\caption{\label{fig:fliptiltillusion_demo} Demonstrating the flip tilt illusion. These images contain homo-pairs of 
dot (each pair has two white or two black dots)  and hetero-pairs of dots (each pair has one black and one white dots). 
The orientation of each dot pair is defined as that of the axis of their alignment. 
All the homo-pairs on the rings in A, B, C, D are parallel to the tangent of the ring.
The dot-pairs on each ring are in the peripheral visual field when viewed while
fixating at the ring's center. Among them, the hetero-pairs (but not homo-pairs) 
should appear --- by the flip tilt illusion ---  tilted orthogonal to their true 
orientation unless we direct our gaze to them to place them in the central visual field.   
Consequently, in noise, conspicuousness of the rings decreases from ring A to ring D.
}
\end{figure}

Figure \ref{fig:fliptiltillusion_demo} demonstrates the flip tilt illusion \citep{ZhaopingFlipTilt2020}, 
using homo- and hetero-pairs of dots tilted parallel or orthogonal to the tangent of the rings.
Ring A contains 100\% homo-pairs tilted tangential to the ring. 
Ring D is like ring A, except all dot pairs a hetero-pairs.
In noise, ring A is highly conspicuous,  but ring D is hard to spot.  
Wherever one directs the gaze on each image, most segments of the ring is in the 
peripheral visual field. 
The flip tilt illusion, normally only visible in the peripheral visual field, makes one perceive 
a hetero-pair of dots as if it is tilted orthogonal to the actual tilt of the dot pair. 
This illusion makes the dot pairs in ring D appear orthogonal to the tangent of the ring, making
ring D inconspicuous in noise.  
(With the illusion, ring A and ring D appear analogous to a snake and a ladder studied by 
May and Hess \citep {MayHess2007}, who showed that ladders but not snakes are hard to spot in the peripheral visual 
field in a noisy background.) Meanwhile, ring B and ring C of Figure \ref{fig:fliptiltillusion_demo} contain
alternating homo-pairs and hetero-pairs. All their homo-pairs are tangential to the ring, but
all their hetero-pairs are orthogonal (in ring B) or parallel (in ring C) to the ring's tangent.
Consequently, due to the illusion, ring B is more conspicuous than ring C in noise, especially when
gaze is not directed to any ring segment. However, when gaze is directed to a ring segment to place
it in the central visual field, the illusion is absent, making this ring segment 
smoother and easier to identify in ring C than in ring B.

The observation that the flip tilt illusion is present in the peripheral but not the 
central visual field was predicted by the central-peripheral dichotomy (CPD) theory \citep {ZhaopingNewFramework2019}.
The theory was motivated by a fundamental constraint that we recognize
only a tiny fraction of retinal inputs because our brain has a processing, or
attentional, bottleneck. Hence, vision has
been formulated as looking and seeing through the bottleneck \citep {ZhaopingNewFramework2019, ZhaopingLookingAndSeeing2026}.  
Looking is to select this fraction for the bottleneck, largely through gaze shift to center
the selected content at the center of gaze. Seeing is to recognize
this content. In this looking-and-seeing framework for vision, the CPD theory states that the 
peripheral visual field is specialize for looking \citep{ZhaopingPeripheral2024}, 
particularly to decide where in the peripheral field should be the
destination for the gaze shift, and that the central visual field is specialized
for seeing. 
The CPD theory predicted the flip tilt illusion in the peripheral visual field because
the theory has an additional hypothesis: the top-down feedback to
aid recognition in light of the bottleneck is mainly focused onto the representation
of the central visual field \citep {ZhaopingNewFramework2019}.
To explain how this prediction arises, we first provide some more background information.

\begin{figure}[hhhh!]
\centering
\includegraphics[width=\textwidth]{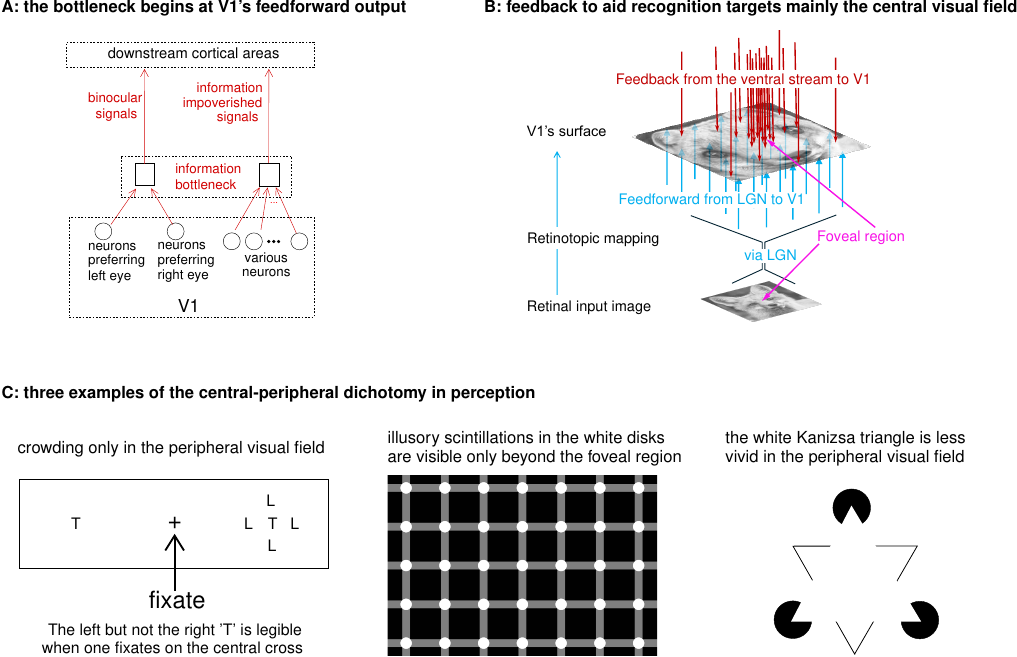}
\caption{\label{fig:FeedforwardFeedback}  
The information bottleneck along the visual pathway (A) and the
central focuses of the feedback to V1 (B) according to 
the central-peripheral dichotomy (CPD) theory.
A: the information bottleneck starts at V1's output to downstream areas, discarding
information such as eye of origin of visual inputs. 
B:  in light of the  information bottleneck, visual recognition relying on impoverished information
downstream from V1 (mainly by the ventral stream of the visual pathway) is difficult.
Feedback, directed mainly to the representation of the central visual field, queries for 
additional relevant information from upstream areas such as V1 to support ongoing recognition.
}
\end{figure}

In the recent decades, converging evidence is supporting the V1 Saliency Hypothesis 
(V1SH) \citep {LiPNAS1999, LiTICS2002} which proposed that the primary visual cortex (V1) supports looking by constructing a saliency map from external inputs 
to guide gaze exogenously to the salient locations (reviewed in \citep{ZhaopingBook2014, ZhaopingV1Function2026}).
Although other brain areas such as the frontal eye field (FEF) and the parietal cortex also control attention and 
gaze shifts through endogenous and goal-dependent factors \citep{BuschmanMiller2007, BisleyGoldberg2010, ZhouDesimone2011}, 
V1's critical role in looking motivated the idea that the bottleneck starts at V1's
output to downstream visual areas along the visual pathway, see  Figure \ref{fig:FeedforwardFeedback}A.
Indeed, by V2, the first cortical area downstream from V1 along the visual pathway, information about the eye of origin
of visual inputs is largely lost, since V1 is the only cortical area with a substantial percentage of neurons that
are monocular to encode eye of origin, which is not discriminable by human observers.
Indeed, a V2 binocular neuron preferring a particular orientation can pool inputs from V1 monocular neurons 
preferring the same orientation but a diversity of preferences for eye of origin.
Future investigations should ascertain how much and what other information are starting
to be lost from the output of V1.

Due to the bottleneck, visual recognition by impoverished information downstream from V1
is difficult.  Accordingly, the CPD theory further hypothesizes that, to aid seeing (recognition),
downstream visual stages send feedback to upstream stages (such as V1) to query for additional information, 
and that this feedback query is mainly directed to the representation of the central visual field \citep {ZhaopingNewFramework2019}.  
Indeed, retrograde tracer injections in marmoset monkeys have confirmed that feedback to V1 from 
recognition-specialized ventral stream of the visual pathway is order of magnitude denser to a unit area of V1 surface
encoding a central rather than a peripheral visual field location \citep {MajkaZhaopingRosa_VSS2026} (Figure \ref{fig:FeedforwardFeedback}B).

The peripheral visual field lacks the feedback query, recognition has to proceed with only the impoverished information 
fed forward through the bottleneck.  Peripheral perception is thus more 
vulnerable to perceptual ambiguity or even perceptual errors, i.e., illusions.
Visual crowding \citep {WhitneyLevi2011}  (Figure \ref{fig:FeedforwardFeedback}C) is one such example of 
visual ambiguity, and the illusory scintillation in the central panel of Figure \ref{fig:FeedforwardFeedback}C 
is not visible wherever one directs the gaze.
The flip tilt illusion  is one of the two novel illusions predicted by the CPD theory for the
peripheral visual field (detailed in the next paragraph).  The other illusion is the reversed 
depth illusion \citep{ZhaopingAckermann2018}, which has also been confirmed as normally only visible
in the peripheral visual field. 

\subsection{The Feedforward-Feedback-Verify-reWeight (FFVW) algorithm for analysis-by-synthesis }

Meanwhile, the illusions are invisible in the central visual field, which uses the additional information
queried by the feedback to veto the initial perceptual hypothesis based on impoverished and misleading information
through the initial feedforward sweep. This perceptual process uses the
Feedforward-Feedback-Verify-reWeight (FFVW) algorithm \citep{ZhaopingFFVW2017} as follows.
In the feedforward step of the FFVW algorithm, the initial feedforward information,
although impoverished by the bottleneck, proposes the initial perceptual hypotheses about the visual scene. 
For example, if perception is trying to decide whether a particular visual object is 
tilted clockwise or counter-clockwise from vertical, the initial hypotheses (or guesses) can be made 
based on the feedforward information.  

However, if the initial feedforward information is insufficient, 
neither of the two perceptual hypotheses can be excluded with sufficient 
certainty, giving ambiguous perception (such as in visual crowding).  For example, 
the available information may propose a 70\% weight and a 30\% weight to the clockwise and counterclockwise tilts.
Without any additional information, the perceptual outcome is most likely the current best guess, i.e., the 
hypothesis (e.g., clockwise tilt) with the highest weight. Illusions can occur, when the initial feedforward 
information is misleading.  However, if the perceptual process can send a feedback to upstream, information-richer, stages, 
such as V1, to query for additional information relevant for discriminating between the alternative hypotheses,
the weights can be re-evaluated. These subsequent steps constitute the Feedback-Verify-reWeight (FVW) components in the FFVW algorithm.
The new weights may make the highest weighted hypothesis even more highly weighted, thereby confirming the initial perception. 
They may re-order the weights, thereby vetoing the initial best guess, which may be an illusion.

The feedback query can iterate, such that, if perception is still too ambiguous
based on the available information from the initial feedforward sweep and
the subsequent feedback queries,  another query for additional information could be sent 
by the feedback. The iteration can terminate when perception is sufficiently unambiguous or when 
the relevant visual input information is no longer available (e.g., by limitation in viewing duration) 
for the query.

A critical ingredient in the FVW components is to query for additional information relevant for discriminating between
the alternative perceptual hypotheses initialized by the feedforward information.
This involves the analysis-by-synthesis computation to identify what would be the responses of the 
upstream neurons if any one of the hypotheses holds.  
This ability to synthesize, or imagine, the would-be responses means that the visual system has 
already acquired an internal model of the visual world. 
During the feedback query, the would-be, synthesized, responses for a given hypothesis can be 
compared with the actual upstream responses, and the weight for this hypothesis can be increases or decreased, 
if the would-be and the actual responses match or otherwise.  

The FVW components can make the white Kanizsa triangle (in Figure \ref{fig:FeedforwardFeedback}C) appear vivid. 
In this case, the fragmentary feedforward information generates the hypothesis for the presence of the
disks and triangles and their relative depth orders, while the feedback query can verify,
e.g.,  the occlusions between the hypothesized object, thereby increasing the weight for the original hypothesis.
Since the feedback query is focused on the central visual field, thus the Kanizsa triangle appears more vivid 
in the central rather than the peripheral visual field. This feedback query is likely the basis for better 
realization in the central visual field of some other related perceptual phenomena  including transparency, 
three-dimensional shape from two-dimensional geometry, and shape from closure, even when the
sizes of visual inputs are appropriately scaled to compensate for acuity differences \citep {MooreEtAl2025}.

\subsection{Prediction of the flip tilt illusion, and of its veto by the feedback query in the central visual field}

\begin{figure}[hhhh!]
\centering
\includegraphics[width=0.5\textwidth]{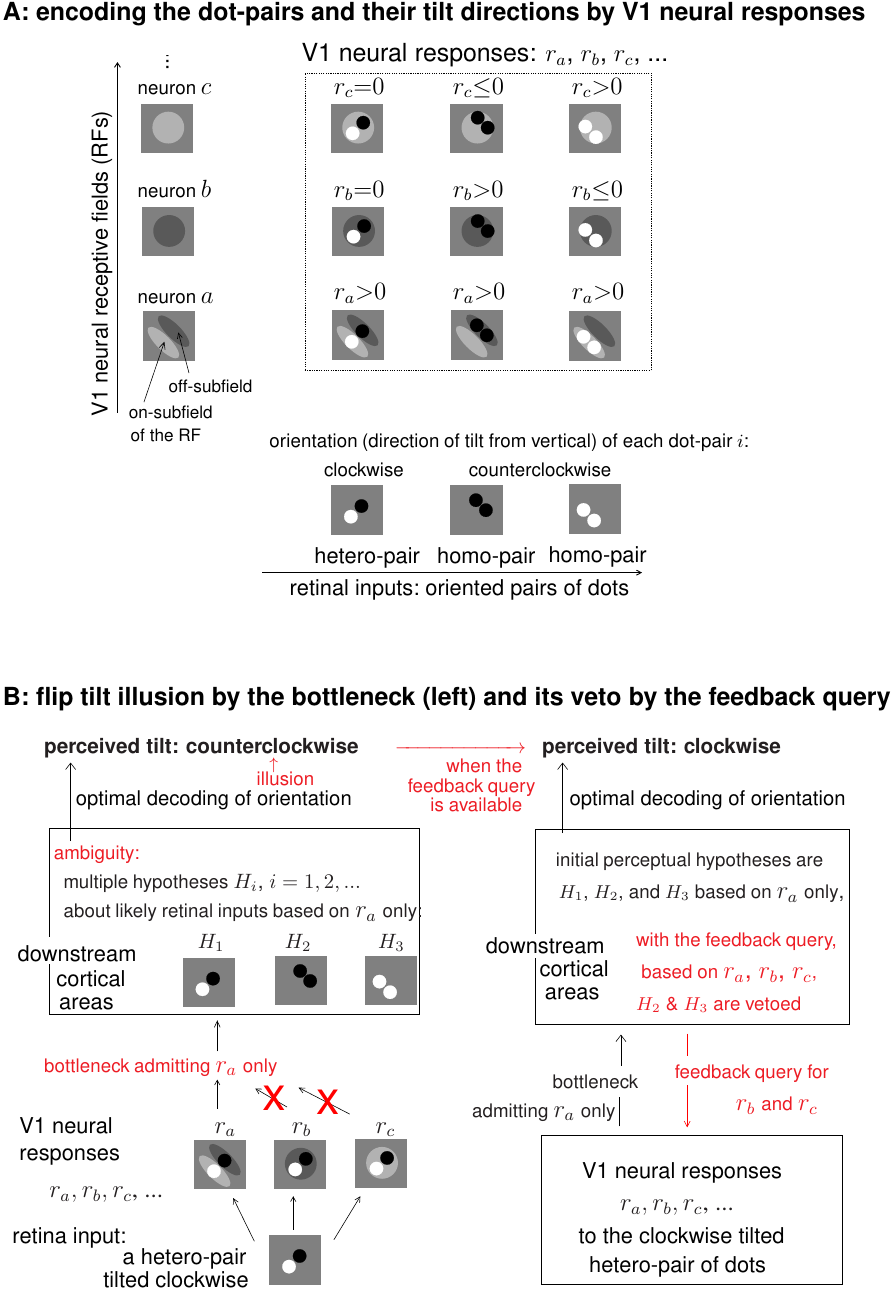}
\caption{\label{fig:fliptiltillusion_derive}  
Derivation of the flip tilt illusion by a bottleneck right after V1, and the veto of this 
illusion by the feedback query.  
A: encoding of retinal inputs of dot-pairs, each denoted as $i$, by V1 neural responses
$(r_a, r_b, r_c, ...)$ resulting from convolving each neural receptive field with the retinal image.
Discrimination between different dot-pairs $i$ is difficult based on
the response $r_a$ alone but is easy by a population response pattern, e.g., $(r_a, r_b, r_c)$. 
B:   the flip tilt illusion is predicted if the bottleneck blocks relevant V1 signals (e.g., neural responses $r_b$ and $r_c$) 
needed to disambiguate between a hetero-pair of dots and homo-pairs of dots orthogonal to the hetero-pair.  
This illusion makes a hetero-pair of dots appear tilted orthogonal to its actual tilt. 
It can be vetoed by feedback to V1 to query for the relevant information, e.g., 
from the responses $r_b$ and $r_c$ of V1 neurons $b$ and $c$.
}
\end{figure}

This paper further investigates the the FFVW algorithm using the flip tilt illusion.
First is a visual psychophysics experiment to test the CPD prediction that the flip tilt illusion should become 
visible in the central visual field by disrupting the feedback. Second is computational modeling to 
reveal the feedforward and feedback components behind the experimental data.
Before we start, Figure \ref{fig:fliptiltillusion_derive} explains how the flip tilt illusion is predicted 
in the peripheral visual field by the CPD theory, and how a feedback query can veto it.

Figure \ref{fig:fliptiltillusion_derive}A shows three dot-pairs --- a hetero-pair tilted clockwise from
vertical, a black homo-pair and a white homo-pair tilted counter-clockwise from vertical --- 
and their evoked neural responses $(r_a, r_b, r_c)$ in three V1 neurons denoted by $a$, $b$, and $c$.
Each dot-pair, including the hetero-pair, 
excites neuron $a$, which prefers a counterclockwise tilt, since the hetero-pair's 
black and white dots fall into this neuron's off- and on-subfields.
Neuron $b$ and $c$ respond selectively to the black and white
homo-pairs.

During recognition, downstream areas entertain  hypothesis $H_1$, $H_2$, or $H_3$
that the retinal input is the hetero-pair, the black homo-pair, or the
white homo-pair, based on V1 responses admitted into the bottleneck.
(For simplicity of argument, we omit other possible retinal inputs.)
If the clockwise tilted hetero-pair is shown but only $r_a$ is admitted into the bottleneck,
perception is ambiguous: all three hypotheses can produce $r_a>0$.
Perceived tilt direction is likewise ambiguous, with $H_2$ and $H_3$ voting
counterclockwise and $H_1$ clockwise,  so a forced-choice
report is likely counterclockwise from the majority voting.
Thus, when the bottleneck admits only $r_a$ via the initial feedforward step of the FFVW algorithm,
a retinal hetero-pair induces a flip tilt illusion
in which perceived orientation is orthogonal to the true orientation.
Formally, substantial conditional probability $p(r_a|H_i)$ of $r_a$
given $H_i$ and prior probability $p(H_i)>0$ for $H_i$ yield
comparable posterior $p(H_i|r_a)$ between $i$'s,
and orientation inference follows
$p(\textrm{orientation} | r_a) =\sum_i p(\textrm{orientation} |H_i) p(H_i|r_a)$.

If the FVW components of the FFVW algorithm is available,
the downstream areas send feedback to V1 to query for $r_b$ and/or $r_c$.
Such feedback requires brain's internal knowledge of the expected or would-be responses 
$r_b$ and $r_c$ for each $H_i$ (i.e., $p(r_b, r_c|H_i)$), i.e., that
$r_b$ and $r_c$ discriminate between the hypotheses.
Comparing the would-be with the queried V1's actual responses $r_b=0$ and $r_c=0$ to the 
hetero-pair veto hypotheses $H_2$ and $H_3$, leaving only $H_1$.
This removes the perceptual ambiguity and vetoes the flip tilt illusion.
Since the central visual field has the privilege of the feedback query (according
to the CPD theory), the flip tilt illusion is normally not visible in the
central field, as demonstrated in Figure \ref{fig:fliptiltillusion_demo}. 

In the derivation above for the flip tilt illusion and its veto, 
we assumed that V1 response $r_a$ was fed forward initially through the bottleneck whereas 
the responses $(r_b, r_c)$ are the feedback queried responses.
However, the exact nature of the initial and subsequently queried information is not critical.
The essential elements include the following.  First, the initial feedforward information is impoverished 
by the bottleneck, so that decoding based on such information can be substantially ambiguous or 
likely erroneous; second, the queried information should contain information not available in the initial 
feedforward sweep and should be discriminative between the alternative perceptual hypotheses 
in the ongoing perceptual ambiguity.

\subsection{The flip tilt illusion is predicted to become visible in the central visual field by backward masking} 

Naturally, the CPD theory predicts additionally that the flip tilt illusion becomes
visible in the central visual field if the feedback query is impaired.
Such a prediction applies also to the reversed depth illusion, for which the prediction 
is confirmed recently \citep{Zhaoping_iScience2025}, using a psychophysical 
trick to disrupt the feedback by backward masking. In backward masking, a visual
stimulus is viewed briefly for a duration $t$ before replaced by a mask.
When $t$ is sufficiently short, the V1 responses available to the 
feedback query will be responses to the mask instead, making
the query ineffective when the mask is uninformative about the original stimulus. 

In the rest of the paper, we  first report an experimental test of this prediction on the flip tilt illusion 
using backward masking.  In anticipation, the prediction was confirmed: observers perceived the tilt of hetero-pairs 
illusorily when $t\lesssim 20$ ms, too short for the feedback query to function effectively.
Then, we constructed a phenomenological model of the FFVW algorithm to explain the data,
revealing that observers' perception is dominately accounted for 
by the feedforward mechanisms or feedback mechanisms for $t< 20$ ms or $t> 30$ ms, respectively.
The paper will end with a discussion on relationships with other works and other visual perceptual phenomena (such as crowding), 
and on the CPD theory and its implications on understanding 
vision.

\section{Methods for the psychophysical experiment to test the prediction}

\begin{figure}[hhhh!]
\centering
\includegraphics[width=0.5\textwidth]{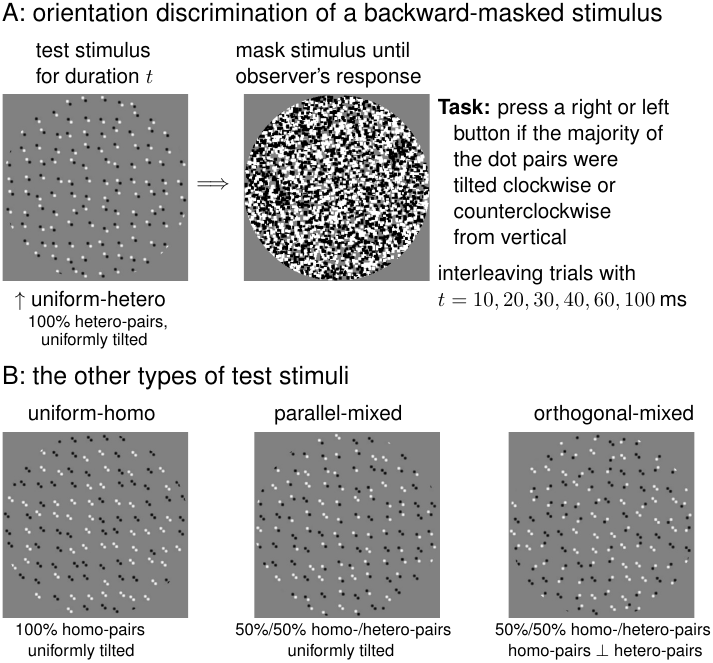}
\caption{\label{fig:Method}  
Experimental design.  A: the essential steps in a testing trial, preceded by a blank screen for $200$ ms, which is in turn preceded by 
a central fixation display for $700$ ms.  Observers presses a button to start the fixation display. 
B: a test session randomly interleaved trials containing test stimuli of types uniform-hetero (see A), uniform-homo, parallel-mixed, and orthogonal-mixed
of various $t$'s.  Observers first practiced for the task on only the illusion-free uniform-homo stimuli in practice sessions.
}
\end{figure}

\subsection{Experimental design}

Figure \ref{fig:Method} outline the experimental design. In each experimental trial, the observer must report whether
dot-pairs in a backward-masked test stimulus were mostly tilted clockwise or counterclockwise from vertical.
The test stimulus was displayed on the center of a SONY cathold ray tube 
(resolution 1232 x 925 pixels, spanning $38.4 \textrm{cm} \times 29.0 \textrm {cm}$, 
and thus $42.0^o  \times 32.3^o$ visual angle at a viewing distance of $50$ cm, 
frame rate was $100$ hz).  
Throughout each experimental session, the gray background in the text stimulus (shown 
in Figure \ref{fig:Method}A) was extended on the whole display. 

Each trial started with a prompt text ``press any button for the next trial" 
in black on the center of the display.  After observer's button press, 
the following four display images appeared one after another:
(1) a fixation display containing a central black fixation cross 
(spanning $5\times 5$ pixels) for $700$ ms, 
(2) a blank display of the gray background for $200$ ms, (3) the test stimulus in 
the center of the display, spanning 370 x 370 pixels 
($11.5 \textrm{cm} \times 11.5 \textrm{cm}$, or $13.1^o \times 13.1^o$ 
in visual angle), for a duration of $t = 10, 20, 30, 40, 60$, or $100$ ms, 
and (4) the mask stimulus spanning 370 x 370 pixels to replace the pixels for the test stimulus.
The mask stayed until the observer pressed a right or left button to report whether most of the dot 
pairs in the test stimulus were tilted clockwise or counterclockwise from vertical. 
Observers were informed that the orientation of a dot pair was defined by that of the 
imaginary line linking the centers of the two dots, and that, 
if different dot-pairs appeared to tilt in different directions, they should report what they 
perceived as the majority direction.  Observers viewed the stimulus naturally 
in their central visual field, but their gaze position was not monitored.

The test stimulus contained homo- and/or hetero-pairs of dots tilted $45^o$ 
clockwise or counterclockwise from vertical.  There were four different types of test stimuli: 
uniform-hetero, uniform-homo, parallel-mixed, and orthogonal-mixed, 
and their examples were shown in Figure \ref{fig:Method}.  
In a uniform-hetero, uniform-homo, or parallel-mixed stimulus, all the dot-pairs were tilted 
uniformly in the same direction.  
In contrast, in a orthogonal-mixed stimulus, the homo-pairs were uniformly tilted 
in a direction orthogonal and opposite to that of the uniformly tilted hetero-pairs.
A uniform-hetero or uniform-homo stimulus contained only hetero-pairs or homo-pairs, respectively.
In parallel-mixed and orthogonal-mixed stimuli,
each dot-pair was equally likely to be a homo-pair or a hetero-pair.  
For each example stimulus in Figure \ref{fig:Method}, another stimulus of the same type 
but with the other tilt direction resembles statistically the mirror reflection of the example 
stimulus about the vertical axis.

When the stimulus type was orthogonal-mixed, the veridical direction of tilt was in principle ambiguous; but
in practice, our $t\le 100$ ms (or even for $t\sim 1$ second) was too short for observers to count the dot-pairs, 
such that they typically reported the tilt direction of the homo-pairs in the stimulus 
(e.g., they would typically report counterclockwise tilt for the orthogonal-mixed stimulus in Figure \ref{fig:Method}B). 
For clarity, this paper refers to the tilt direction of such a stimulus also by that of the homo-pairs.

If the CPD prediction holds, observers should see the flip tilt illusion in a uniform-hetero stimulus 
when $t$ was too short to allow the feedback query to function effectively.
Accordingly, their task performance on such a stimulus should be better or worse than the chance level 
when $t$ is sufficiently long or sufficiently short, respectively.
The uniform-homo is a control stimulus to verify that, for illusion-free homo-pairs, 
small $t$ does not depress the task performance below the chance level.
The parallel-mixed and orthogonal-mixed stimuli provide additional verifications and insights detailed later.

To ensure that observers could do the task sufficiently well for small $t$, 
observers were first trained in practice sessions using only the uniform-homo stimuli (detailed below), 
before proceeding to test sessions which randomly interleaved all the four types of the stimuli.


\subsection{Details of the Stimuli}

A practice session contained $270$ trials of uniform-homo stimuli and took
about 15 minutes (including a very brief break after half of the trials). 
It randomly interleaved trials of six different stimulus 
durations $t = 10, 20, 30, 40, 60, 100$ ms, 50 trials for each $t<100$ ms and 20 trials for $t = 100$ ms.
A test session contained $n_{\textrm{conditions}} = 24$ possible stimulus conditions, 
each defined by a particular combination of the stimulus type (uniform-homo, uniform-hetero, parallel-mixed, or orthogonal-mixed)
and $t = 10, 20, 30, 40, 60$, or$100$ ms.  Each condition had $n_{\textrm{trials per condition}} = 14$ or $6$ trials when $t < 100$ ms or $t = 100$ ms, 
with $n_{\textrm{trials per condition}}/2$ trials for each tilt direction.
A test session had $304$ trials and took about about 15 minutes to complete, including a brief break in the middle.

Each test stimulus was made by cropping a square image containing $370 \textrm{~pixels} \times 370 \textrm{~pixels}$ (spanning $13.1^o \times 13.1^o$  in visual angle), 
such that all the pixels outside the central circular area (of $370$-pixel diameter) had the background gray level.
The original square image had $12\times 12$ dot-pairs distributed in 
a regular $12\times 12$ regular grid ($30.8 \textrm{pixels} \times 30.8 \textrm{pixels}$ per unit grid). 
From its corresponding grid position, each dot-pair was positioned randomly and independently within a $\pm 7.8$ pixels horizontally and vertically.

Given the condition and tilt direction of a stimulus,
there were only $n_{\textrm{dot-pairs}} = 2$ or $n_{\textrm{dot-pairs}} =4$  possible kinds of dot-pairs 
in the stimulus.  
For example, a counterclockwise-tilted, uniform-hetero, stimulus contains $n_{\textrm{dot-pairs}} = 2$ 
possible kinds of counterclockwise-tilted hetero-pairs, one with the black dot above the white dot, 
and the other with the the white dot above the black dot (see Figure \ref{fig:Method}A).
Meanwhile, an orthogonal-mixed stimulus with counterclockwise-tilted homo-pairs (see Figure \ref{fig:Method}B) 
has $n_{\textrm{dot-pairs}} =4$ possible kinds of dot-pairs:  counterclockwise-tilted white homo-pair, counterclockwise-tilted black homo-pair, 
clockwise-tilted hetero-pair with a black dot above the white dot, and clockwise-tilted hetero-pair with a white dot above the black dot.
Accordingly, each of the $12\times 12$ dot pairs was randomly  and equally likely to be one of 
the $n_{\textrm{dot-pairs}}$ possible kinds of dot-pairs for a given stimulus condition and tilt direction. 

Each black or white dot in a dot-pair was depicted by image pixels whose luminance deviated from that of the gray background according to a 
gaussian profile $\propto \exp(-r^2/(2\sigma ^2))$, in which $r$ is the distance between the pixel and the center of the dot 
and $\sigma = 2.3$ pixels.  The centers of the two dots in a dot-pair was $8$ pixels apart from each other.
The random choice of position jitters and the random choice of the dot-pair types were independent 
between dot-pairs and between trials.

Each mask stimulus was made by randomly placing black and white small squares (each with side length $6$ pixels) on the gray background.
The density of these small squares was $40$ times of the density of the dot-pairs in a test stimulus. The neighboring random squares
could occlude each other. Like the test stimulus, the image area for the masking squares was cropped to within the 
circular area, and the masks in different trials were generated independently and were independent of any stimuli.

\subsection{Experimental participants}

Before any data taking, each observer was shown at least two examples (for the two possible tilt directions) of each stimulus type 
in Figure \ref{fig:Method}.  Each example was freely viewed (usually for up to several seconds)  for the observer to report the 
tilt direction for the example.  Each observer gave the correct or expected report in all or most such stimulus demonstration trials.
If an incorrect or unexpected report was given, more stimulus examples, particularly of orthogonal-mixed stimuli, would be shown to 
make sure that the task was understood correctly.  Observers were informed about the following:  in real data taking sessions, each test stimulus 
would be viewed only for a very brief duration, the correct tilt direction was equally likely to be clockwise or counterclockwise,
they should try their best (as if in a competition) to report the correct direction, guess if they had to.

Each observer started data taking in practice sessions containing only uniform-homo stimuli.
Their performance typically improved over successive practice sessions, such that,
after a few (typically two or three) practice sessions, it typically reached
a qualified performance level, which is defined by nearly perfect 
performance for $t\ge 40$ ms and substantially better than chance performance for $t= 20$ ms.  
Fifteen out of twenty one naive observers consistently achieved qualified 
performance in practice sessions, and all of them were admitted to proceed to test sessions.   

Once admitted to test sessions, each of the $n=15$ observers took seven to ten test sessions, 
yielding a total of $2128$ to $3040$ trials. Each observer had no more than four sessions per day and had 
sufficient rests between sessions.  Whether an observer took more than seven test sessions was determined by 
the convenience of the observer and session scheduling.  The observers were not informed that 
the test stimuli in the earlier (practice) sessions contained only homo-pairs of dots.
There were aware that different trials had different stimulus viewing durations $t$, and noticed 
that trials with shorter $t$'s  were very difficult. Each observer was often reminded, particularly before 
the first testing or practicing session of a data-taking day, of all the four types of stimuli
and of the task by showing them some example stimuli for sufficiently long $t$.
The results reported in this paper were obtained using data from the first seven test sessions of each observer, 
so that each observer contributed the same number of trials to the data analysis.
The qualitative conclusions of this paper are unchanged if each observer contributed all their testing trials instead.

All observers had normal or corrected-to-normal vision. The study was conducted according to the Declaration of Helsinki (1964), and ethical approval was 
obtained from the University Clinic T\"ubingen ethics committee.

\subsection{Statistical methods}

For each stimulus condition, if an observer $o$ pressed the correct button in $n$ out of a total $N$ trials accumulated over all their testing 
sessions, their  accuracy is $p_o = n/N$.  
Let $A$  and $\Sigma$ denote the mean and standard deviation of $p_o$ across the $N_{\textrm{observers}} = 15$ observers, then
 $A = \sum_o p_o/N_{\textrm{observers}}$, and the standard error of this mean is $\epsilon = \Sigma /\sqrt{N_{\textrm{observers}}}$
in typical conditions. The condition of uniform-homo stimulus with $t = 100 $ ms was exceptional by having $p_o$ for all observers.
In this condition, the conventional estimations of $A$ and $\epsilon$ are replaced by the Bayesian estimations, 
given a total of $N=630$ trials of this condition accumulated across the $N_{\textrm{subjects}} = 15$ subjects, 
of $A= (N+1)/(N+2) =0.9984$ and $\epsilon  = \sqrt{ (N+1)/((N+2)^2 (N+3))} =0.0016$.

To test whether the $A$ for two different conditions were significantly different from each other 
paired two-tailed 
t-test was used on $p_o$ across observers for the two 
conditions concerned.   Sometimes, paired one-tailed t-test was used, this will be indicated in the text.
Similarly, whether $A$ is significantly different from the chance level of $A=0.5$ was ascertained by two-tailed t-test.

\begin{figure}[hhhh!]
\centering
\includegraphics[width=0.7\textwidth]{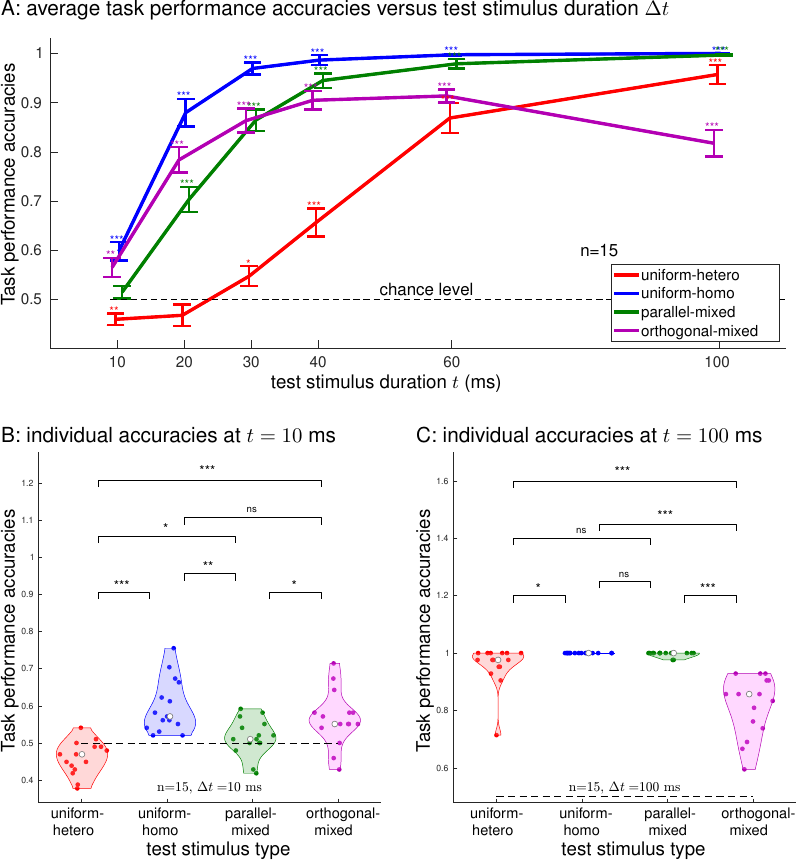}
\caption{\label{fig:ResultSummary1}
Summary of experimental results.  A: mean task accuracies $A$ in the test sessions, averaged across observers.
The error bars mark the standard errors $\epsilon$ of the mean. The accuracies that are significantly
different (by two-tail t-test) from the chance level of $0.5$ are indicated by `*' (for p-value $0.01\le p< 0.05$), 
`**' ($p<0.01$), or `***' ($p<0.001$) above the error bars.    
At $t = 10$ ms, the mean accuracy $A$ on the uniform-hetero stimuli was 
significantly below the chance level ($p=0.004$), manifesting the flip tilt illusion. 
At $t = 20$ ms, this accuracy was still below chance, but insignificantly.
B: accuracies of the individual observers at $t = 10$ ms are marked as colored dots in a 
violin plot, the white dots mark the median accuracies across observers. 
Outcomes of statistical tests (paired two-tailed t-test) between the mean accuracies of 
two different conditions are indicated (by 'ns' for no significant differences or by stars for significant differences) 
above the horizontal brackets linking the stimulus types being compared.
C: as B, but for $t = 100$ ms.
}
\end{figure}

\section{Results from the psychophysical experiment}

\begin{figure}[hhhh!]
\centering
\includegraphics[width=0.7\textwidth]{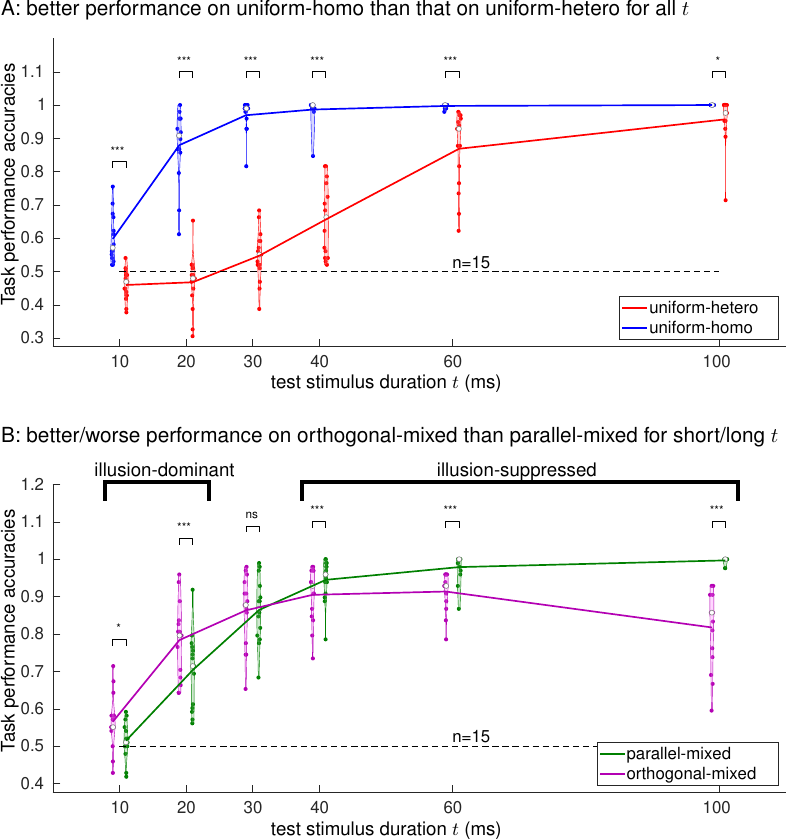}
\caption{\label{fig:ResultSummary2}
Comparing accuracies on uniform-homo and uniform-hetero (A), 
and accuracies on parallel-mixed and orthogonal-mixed (B).
A: uniform-homo yielded higher accuracies than uniform-hetero for every $t$.
The difference between their accuracies was largest at intermediate $t \sim 30$ ms. 
B: the two accuracy curves for the parallel-mixed and orthogonal-mixed stimuli crossed each other 
at  $t=30$ ms, which separates the illusion-dominant zone from the illusion-suppressed zone.
The performance was significantly better on the parallel-mixed for long $t$,
but was significantly better on the orthogonal-mixed for short $t$ ($p = 0.0125$ at $t=10$ ms, one-tailed t-test).
At $t = 30$ ms, the two accuracies are indifferent from each other ($p=0.93$, two-tailed t-test).
In both A and B, 'ns', '*', '**', and '***' indicate that $p\ge 0.05$, $0.01\le p< 0.05$, 
$p<0.01$, and $p<0.001$ by the paired two-tailed t-tests comparing the two accuracies at each $t$.
}
\end{figure}

The chance level accuracy is $0.5$ for random button presses in this experiment.
Averaged across observers, only uniform-hetero stimuli evoked accuracies
below the chance level at small $t \le 20$ ms, significantly so for $t = 10$ ms 
(p-value $p=0.004$ by two-tailed t-test), see Figure \ref{fig:ResultSummary1}A.
Hence, the flip tilt illusion was visible in the central visual field when the feedback query was 
interrupted by backward masking, as predicted by the CPD theory.

At $t = 10$ ms, observers were near the limit of human visual temporal resolution.
Indeed, their mean accuracy on uniform-homo stimuli was around 0.6, 
approaching, but significantly higher than, the chance level ($p <0.001$).
Presumably, the feedback query is ineffective at this $t$. 
This is consistent with accuracies on parallel-mixed and orthogonal-mixed stimuli.
In orthogonal-mixed stimuli, both the homo-pairs and the hetero-pairs 
excite V1 neurons preferring the orientation of the homo-pairs, such that their V1's orientation signals
reinforce each other, see Figure \ref{fig:fliptiltillusion_derive}. 
Indeed, the accuracy on orthogonal-mixed stimuli was statistically equivalent ($p=0.08$) to that on the 
uniform-homo stimuli, and was significantly ($p=0.004$) above the chance level (see Figure \ref{fig:ResultSummary1}AB).  
In contrast, in response to parallel-mixed stimuli,  V1's orientation signals 
evoked by the hetero-pairs oppose those evoked by the homo-pairs.
Indeed, the corresponding accuracy was indifferent from the chance level ($p=0.261$).

At $t  = 100$ ms (Figure \ref{fig:ResultSummary1}C), the feedback 
query should be effective to veto the flip tilt illusion.
Indeed, the accuracy for the uniform-hetero stimuli was nearly perfect, 
although statistically lower than that for the uniform-homo stimuli ($p=0.04$, two-tailed t-test).  
Similarly, the accuracies for the uniform-homo and parallel-mixed are statistically equivalent ($p= 0.16$). 
In contrast, the accuracy of the orthogonal-mixed was significantly lower than those of the other 
stimulus types ($p< 0.001$), since the homo-pairs and hetero-pairs have opposing veridical orientation signals,

Uniform-homo stimuli consistently yielded higher accuracies than uniform-hetero stimuli, particularly at 
intermediate $t$ (see Figure \ref{fig:ResultSummary2}). 
In contrast, accuracy curves for parallel-mixed and orthogonal-mixed stimuli cross each other
at intermediate $t=30$ ms, close to $t=20$ ms at which the accuracy on uniform-hetero stimuli was at the chance level. 
Short $t<30$ ms constitute the illusion-dominant zone, in which the  illusory (flip) tilt dominates 
the veridical tilt so that task performance is better for the orthogonal-mixed than the parallel-mixed stimuli.
Long $t>30$ ms constitute the illusion-suppressed zone, in which the  illusory (flip) tilt is suppressed by the veridical tilt, 
so that the task accuracy is better for the parallel-mixed than the orthogonal-mixed stimuli.

\FloatBarrier

\section{A model to dissect the feedforward and feedback contributions to behavioral performance}

A computational model of the feedforward and feedback components in FFVW is built to explain the psychophysical data.
Let $A_i(t) $ denote the behavioral accuracy in stimulus type
$i = \textrm{homo}$, $\textrm{hetero}$, $\textrm{parallel-mixed}$, or $\textrm{orthogonal-mixed}$ at stimulus 
duration $t$, and $S_i (t) $ be the neural decision signal, 
such that $A_i(t)$  is a sigmoid function of $S_i (t) $, i.e.,
\begin{align}
     A_i (t) &= \sigma (S_i (t)) \equiv  {1 \over {[1+\exp (-S_i (t))]}}, \nonumber \\
 &  \mbox{for $i= \textrm{homo}, \textrm{hetero}, \textrm{parallel-mixed}, \textrm{or orthogonal-mixed}$}.  \label{eq:AccuracyDefinition}
\end{align}
For simplicity, the notations ``uniform-homo" and ``uniform-hetero" are also simply ``homo" and ``hetero".
The $S_i (t) $ includes the feedforward component 
$S_{\textrm{ff}, i}(t)$
and feedback component
$S_{\textrm{fb}, i}(t)$
\begin{align}
   S_{i} (t) =&  S_{\textrm{ff}, i}(t)+
S_{\textrm{fb}, i}(t) \mbox{~~for each $i$}.    \label{eq:SFFandFBDefinition}
\end{align}
The feedforward component
$ S_{\textrm{ff}, i}(t)$ 
arises  from V1's feedforward signals admitted into the bottleneck, as depicted in Figure \ref{fig:fliptiltillusion_derive}.
Meanwhile,  the feedback component
$ S_{\textrm{fb}, i}(t)$ 
arises from the FVW components of the FFVW algorithm. It is associated with recognizing the 
pairs of dots as objects, each with an orientation defined by the implicit line linking the centers of the two dots.
Furthermore, for parsimony, we model the decision signals for the parallel-mixed and orthogonal-mixed stimuli as arising 
from mixing the decision signals of the uniform-homo and uniform-hetero stimuli, 
\begin{align}
   S_{\textrm{parallel-mixed}}(t)  &= \left [S_{\textrm{homo}}(t) + S_{\textrm{hetero}}(t)\right ]/2, \label{eq:Sparallel-mixedDefinition} \\
   S_{\textrm{orthogonal-mixed}}(t)  &=\left  [S_{\textrm{homo}}(t) - S_{\textrm{hetero}}(t)\right ]/2.   \label{eq:Sorthogonal-mixedDefinition} 
\end{align} 
The equations \ref{eq:AccuracyDefinition}-\ref {eq:Sorthogonal-mixedDefinition} above are parameter-free part of our model formulation.
According to this part of the formulation,  accuracies $ A_i (t)$ for all conditions are completely determined once we determine 
$S_{\textrm{homo}}(t)$ and $S_{\textrm{hetero}}(t)$, which we model next phenomenologically. 

The feedforward component $S_{\textrm{ff}, i}(t )$
is modelled as 
\begin{align}
   S_{\textrm{ff}, i}(t ) = & S_{\textrm{ff}, i}(\infty )\cdot \left [1- \exp \left (- t/ T_{\textrm{ff}} \right ) \right ],  \nonumber \\ 
 &  \mbox{for $i=$ homo or hetero.} \label{eq:FF} 
\end{align} 
This equation has three parameters: 
 $S_{\textrm{ff, homo}}(\infty )$,
 $S_{\textrm{ff, hetero}}(\infty )$,
and $T_{\textrm{ff}}$. 
The first two parameterize the maximum possible, or saturation, 
values for $ S_{\textrm{ff}, i}(t)$ 
when $t \rightarrow \infty$.
They model the effect of the feedforward V1 responses on the two-alternative-forced-choice (2AFC) perceptual 
decision task.  Their magnitudes increase with the overall strengths of V1 responses (e.g., 
$r_a$ in Figure \ref{fig:fliptiltillusion_derive} to homo-pairs and hetero-pairs)
in the initial feedforward sweep of information admitted into the bottleneck for the perceptual decision process.  
Meanwhile, 
the sign of 
$S_{\textrm{ff}, i}(\infty )$,
 positive or negative, indicates whether the preferred orientation of the activated neurons (whose responses are admitted into the bottleneck), 
e.g., neuron $a$ with response $r_a$, 
agrees with, or opposes,  the veridical orientation of the dot-pair for the 2AFC perceptual decision.
Hence, 
$S_{\textrm{ff, homo}}(\infty ) >0$ 
and $S_{\textrm{ff, hetero}}(\infty ) < 0$ should hold 
to reflect the fact that V1 neurons most activated by the hetero-pairs (but not homo-pairs)  
prefer an orientation orthogonal to the actual orientation of the hetero-pairs \citep {SmithEtAl2002}.
The parameter $T_{\textrm{ff}}$ is the time scale for the evolution of the feedforward decision signals based on the feedforward V1 responses 
admitted into the bottleneck.  Accordingly, the feedforward decision signal $S_{\textrm{ff}, i}(t )$
grows approximately linearly with time for short time $t \ll T_{\textrm{ff}}$, but saturates in 
time when $t \gg T_{\textrm{ff}}$, like in the empirical Bloch's law for temporal integration 
of sensory signals \citep{Gorea2015}. The process for transforming
initial feedforward responses to decision signals is assumed to be sufficiently simple and straight-forward 
such that $T_{\textrm{ff}}$ takes the same value for both homo- and hetero-pair stimuli.

The feedback component is modelled as, for $i = \textrm{homo}$ or $\textrm{hetero}$, 
\begin{equation}
   S_{\textrm{fb},i}(t ) = 
					H(t - T_{\textrm{ff}}) \cdot
		 			 S_{\textrm{fb}}(\infty ) \cdot   \left [1- \exp \left (-  (t /T_{\textrm{fb}, i} )^2\right )\right ].
				 \label{eq:FB} 
\end{equation} 
Here, $H(t - T_{\textrm{ff}})$ models a step function (with $H(t  - T_{\textrm{ff}})=1$ or $0$ if $t\gtrsim T_{\textrm{ff}}$ or otherwise)
to make the feedback signal $S_{\textrm{fb},i}(t ) \approx  0$ negligible 
when the stimulus duration $t < T_{\textrm{ff}}$ is too short to substantially 
realize the initial feedforward sweep.  When $t > T_{\textrm{ff}} $ holds,
$S_{\textrm{fb},i}(t ) \simeq  S_{\textrm{fb}}(\infty ) \cdot   \left [1- \exp \left (-  (t /T_{\textrm{fb}, i} )^2\right )\right ]$
models the growth of 
$S_{\textrm{fb},i}(t )$ with time $t$, by a time scale 
parameterized by $T_{\textrm{fb}, i}$, towards a saturating magnitude 
$ S_{\textrm{fb}}(\infty ) $ if $t \gg T_{\textrm{fb}, i}$.
In the exponential $\exp \left (-  (t /T_{\textrm{fb}, i} )^2\right )$,
 the argument is quadratic in ($t /T_{\textrm{fb}, i}$).
In comparison, in the  exponential $\exp \left (- t/ T_{\textrm{ff}} \right ) $ for modeling
$S_{\textrm{ff},i}(t)$,
the argument is linear in $(t/ T_{\textrm{ff}})$.
Hence,  the feedback is modeled as a more nonlinear process than the feedforward process in the FFVW algorithm.

Including both $i =$ homo and $i=$ hetero, equation \ref{eq:FB} contains three parameters: 
$S_{\textrm{fb}}(\infty )$,  $T_{\textrm{fb, homo}}$, and $T_{\textrm{fb, hetero}}$.  
The $S_{\textrm{fb}}(\infty )$ parameterizes the effect of completing
the FVW components in the FFVW algorithm, i.e., the effect of recognizing the orientation 
of a pair of dots, regardless of whether the dot-pair is a homo- or a hetero-pair. 
Meanwhile, $T_{\textrm{fb, homo}} \ne T_{\textrm{fb, hetero}}$ by default.
For the homo-pairs,  the FVW process merely confirms the initial (dominant) perceptual hypothesis 
about the tilt direction of the dot-pairs based on the initial feedforward signals. 
In contrast, for the hetero-pairs, the FVW process vetoes the initially dominant hypothesis (which would 
lead to the flip tilt illusion) and re-orders the relative weights for the two competing 
hypotheses (clockwise and counterclockwise tilt directions), see Figure \ref{fig:fliptiltillusion_derive}.

\begin{figure}
\begin{center}
\setlength{\unitlength}{1.13pt}
\begin{picture}(400, 430)
\put(0, 0) {\includegraphics[width=0.86\textwidth]{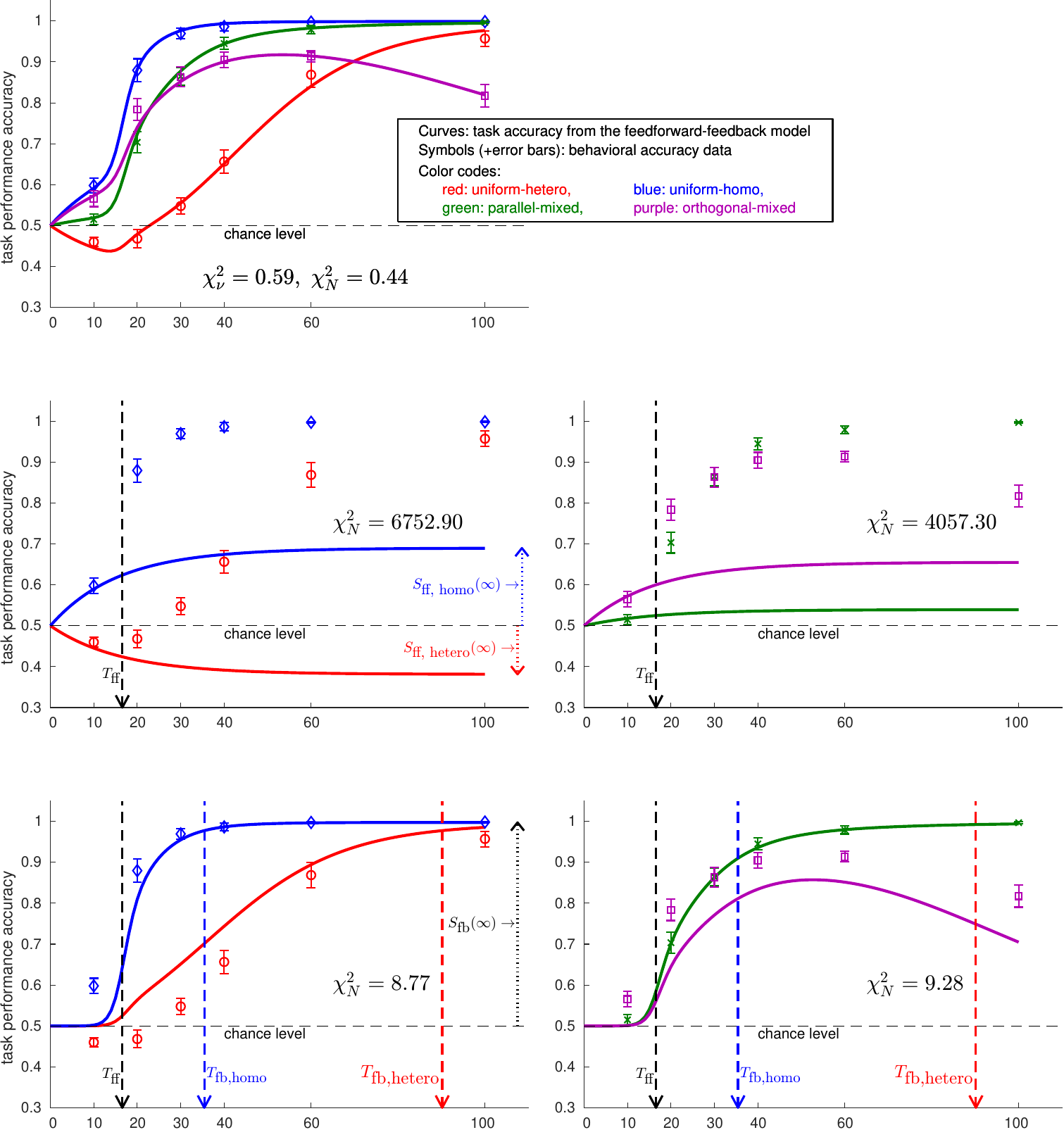}}
\put( 72, -6){\sf\scriptsize Stimulus duration $t$ (ms)}
\put( 262, -6){\sf\scriptsize Stimulus duration $t$ (ms)}
\put( 72, 133){\sf\scriptsize Stimulus duration $t$ (ms)}
\put( 262, 133){\sf\scriptsize Stimulus duration $t$ (ms)}
\put(2, 115){\sf\textbf{C: when the feedforward component in the model in A is removed }}
\put(2, 255){\sf\textbf{B: when the feedback component in the model in A is removed }}
\put(2, 395){\sf\textbf{A: task performance accuracy by the model (curves) and by behavioral data}}
\put( 72, 270){\sf\scriptsize Stimulus duration $t$ (ms)}
\end{picture}
\end{center}
\caption{\label{fig:model}
The model, described by equations \ref{eq:AccuracyDefinition} -\ref{eq:FB}, agrees with the behavioral data on four curves of 
task accuracy versus stimulus duration $t$ for four stimulus types $i$ 
($i=$ uniform-homo, uniform-hetero, parallel-mixed, and orthogonal mixed) for a total $N=24$ data points, 
using a total of $m=6$ parameters (in equation \ref{eq:model_parameters}):  
$S_{\textrm{ff, homo}}(\infty)$,
$S_{\textrm{ff, hetero}}(\infty)$,
$S_{\textrm{fb}}(\infty)$, 
$T_{\textrm{ff}}$, 
$T_{\textrm{fb, homo}}$,
$T_{\textrm{fb, hetero}}$.
A: in most conditions, behavioral accuracy $A_i(t |\textrm{data})$ deviates from the model's 
prediction $A_i(t |\textrm{model})$ by no more than the standard error $\epsilon_i (t|\textrm{data})$ (indicated
by the error bars) of  $A_i(t |\textrm{data})$, with
a reduced chi-square statistics $\chi^2_\nu  = 0.59$.  
B or C: dissecting model behavior by removing the feedback 
(i.e., setting $S_{\textrm{fb}} (\infty)=0$) 
or feedforward 
(i.e., setting $S_{\textrm{ff}, homo} (\infty)=S_{\textrm{ff}, hetero} (\infty) =0$) 
component of the model in A.
The left panels focus on uniform-homo and uniform-hetero conditions, whereas the 
right panels on parallel- and orthogonal-mixed conditions.
The $\chi^2_{\scriptscriptstyle N}$ (see equation \ref{eq:chi-square-N}) in each plot of 
B and C is for the $N=12$ data points in the plot.  
Comparing B and C, one can see that the behavior performance is dominated by the 
feedback mechanisms except for the smallest $t=10$ ms.  
Indeed, $\chi^2_{\scriptscriptstyle N} > 4000$ in B and $\chi^2_{\scriptscriptstyle N} < 10$ in C. 
The black, blue, and red vertical dashed-line arrows point to 
the values of temporal parameters $T_{\textrm{ff}}$, $T_{\textrm{fb,homo}}$, 
and $T_{\textrm{fb,hetero}}$ on the x-axis. 
In the left panel of B and C, the length and direction of each vertical dotted-line arrow 
(let us denote it as $y$) mark the deviation from the chance level of model's asymptotic accuracy
when $t\rightarrow \infty$ using feedforward (B) or feedback (C) mechanisms alone.  
By $y={1\over 2} \tanh (x/2)$,  this $y$ reflects $x$ when $x$ is the magnitude and sign of 
$S_{\textrm{ff, homo}}(\infty)$, $S_{\textrm{ff, hetero}}(\infty)$, or $S_{\textrm{fb}}(\infty)$ 
as indicated.
The left panel in B indicates that  
$|S_{\textrm{ff, homo}}(\infty )| > |S_{\textrm{ff, hetero}}(\infty )|$, consistent
with  V1 neurons' higher sensitivities to the homo-pairs than the hetero-pairs \citep {SmithEtAl2002}. 
B and C indicate that $T_{\textrm{ff}} < T_{\textrm{fb,homo}} < T_{\textrm{fb,hetero}}$, as expected.
The right panels in B and C suggest that perception for parallel-mixed stimuli is
approximately driven by the feedback mechanisms alone.
}
\end{figure}

In any case, a total of six parameters
$S_{\textrm{ff, homo}}(\infty )$,
$S_{\textrm{ff, hetero}}(\infty )$,
$S_{\textrm{fb}}(\infty )$,
$ T_{\textrm{ff}}$, $T_{\textrm{fb, homo}}$, and $T_{\textrm{fb, hetero}}$
are used to model both the feedforward and feedback components in the FFVW algorithm responsible
for a total of $24  = 4\times 6$ data points of the psychophysical accuracy $A_i(t)$ 
for the four stimulus conditions $i= \textrm{homo}$, $\textrm{hetero}$, $\textrm{parallel-mixed}$, and $\textrm{orthogonal-mixed}$
and six $t$'s. Let $A_i(t| \textrm{data} )$ be the mean accuracy across our observers for each 
$(i, t)$, $\epsilon_i (t | \textrm{data})$ be its standard error,
and $A_i(t | \textrm{model})$ be the model's predicted accuracy given a set of
these six parameters. The best parameter set is identified using matlab's optimization routines as the 
one that minimizes the chi-squared statistics
\begin{equation}
	\chi^2 = \sum_{i, t} {\left [A_i(t |\textrm{data}) - A_i(t |\textrm{model})\right ]^2\over \left [\epsilon_i (t |\textrm{data}) \right ]^2}.
	\label{eq:chi-square} 
\end{equation}
A good model makes the following statistics small relative to unity,
\begin{align}
  	\quad \chi^2_{\scriptscriptstyle N} &= { \chi^2 \over N}, \quad \mbox{N = the number of data points,} \label{eq:chi-square-N}\\
  	\quad \chi^2_\nu &= { \chi^2 \over \nu}, \quad \mbox{$\nu = N-m$ is the degree of freedom}, \nonumber \\
		& \quad\quad \mbox{and $m$ is the number of model parameters.} \label{eq:chi-square-reduced}
\end{align}
The optimized model parameters are
\begin{equation}
\begin{aligned}
S_{\textrm{ff, homo}}(\infty)  &= 0.800,  \\
S_{\textrm{ff, hetero}}(\infty) &=-0.484   \\
S_{\textrm{fb}}(\infty ) &= 5.97, \\
 T_{\textrm{ff}} & = 16.6 ~~\textrm{millisecond}, \\
T_{\textrm{fb, homo}} &= 35.4 ~~\textrm{millisecond}, \\
T_{\textrm{fb, hetero}} & = 90.2  ~~\textrm{millisecond}.
\end{aligned} \label{eq:model_parameters}
\end{equation}
They yield $\chi^2_\nu = 0.59$, such that for most conditions $(i, t)$,
we have $|A_i(t |\textrm{data}) - A_i(t |\textrm{model})| < \epsilon_i (t|\textrm{data})$, 
as evident in Figure \ref{fig:model}.
This $\chi^2_\nu$ was obtained 
when the step function $H(t - T_{\textrm{ff}})$ was approximated by 
\begin{equation}
H(t - T_{\textrm{ff}}) \approx 0.5\left [1+\tanh \left( {{t - T_{\textrm{ff}}}\over {T_{\textrm{ff}}/D}}\right ) \right ], \quad
\mbox{with $D = 5$}. 
\end{equation}
As long as $D \gg 1$, the model performance and optimal parameters are not sensitive to the 
exact $D$. For example, replacing  $D = 5$ by  $D=100$ (to make the
step change much more quickly) alters $\chi^2_\nu$ by about $1\%$, 
and alters the optimized value of each model parameter by about 7\% on average and no more than 18\% in each.
Therefore, the quantitative values of the model parameters in equation \ref{eq:model_parameters} should meaningfully 
reflect the properties of the FFVW mechanisms.

The model reveals the following salient features of the feedforward and feedback processes 
in the FFVW algorithm.
\begin{enumerate}
\item Except for very small $t$, perception is dominated by the feedback 
rather than feedforward processes.
This is reflected by 
$S_{\textrm{fb}}(\infty ) \gg S_{\textrm{ff, homo}}(\infty) $
and
$S_{\textrm{fb}}(\infty ) \gg S_{\textrm{ff, hetero}}(\infty) $,
and by the large 
$\chi^2_{\scriptscriptstyle N} > 4000$ if only feedforward signals were used for perception 
(see Figure \ref{fig:model}B) and the much smaller $\chi^2_{\scriptscriptstyle N} < 10$ if only 
feedback signals were used (see Figure \ref{fig:model}C). 

\item  $T_{\textrm{fb,homo}} $ is about twice the value of $T_{\textrm{ff}}$, suggesting that
	confirming the feedforward hypothesis for the homo-pairs may involve just about one iteration 
of the feedback query in the FFVW algorithm.

\item  $T_{\textrm{fb,hetero}}$ is much longer than 
  $T_{\textrm{fb,homo}}$, suggesting that, for hetero-pairs, vetoing a perceptual hypothesis initiated 
by the feedforward sweep involves more complex feedback processing than simply  confirming an 
initial hypothesis for the homo-pairs.  It is likely that  this vetoing involves multiple 
iterations of the feedback query in the FFVW algorithm. 

\item  The values 
$|S_{\textrm{ff,homo}}(\infty)|$
and
$|S_{\textrm{ff,hetero}}(\infty)|$
are too weak such that the 
decision signals $S_{\textrm{ff},i} (t )$  
by feedforward mechanisms are insufficient 
for making perception confident (so that the accuracy approaches $1$ or $0$) even when $t \rightarrow \infty$. 
\end{enumerate}

\begin{figure}
\begin{center}
\setlength{\unitlength}{1.00pt}
\begin{picture}(400, 200)
\put(0, 0) {\includegraphics[width=0.77\textwidth]{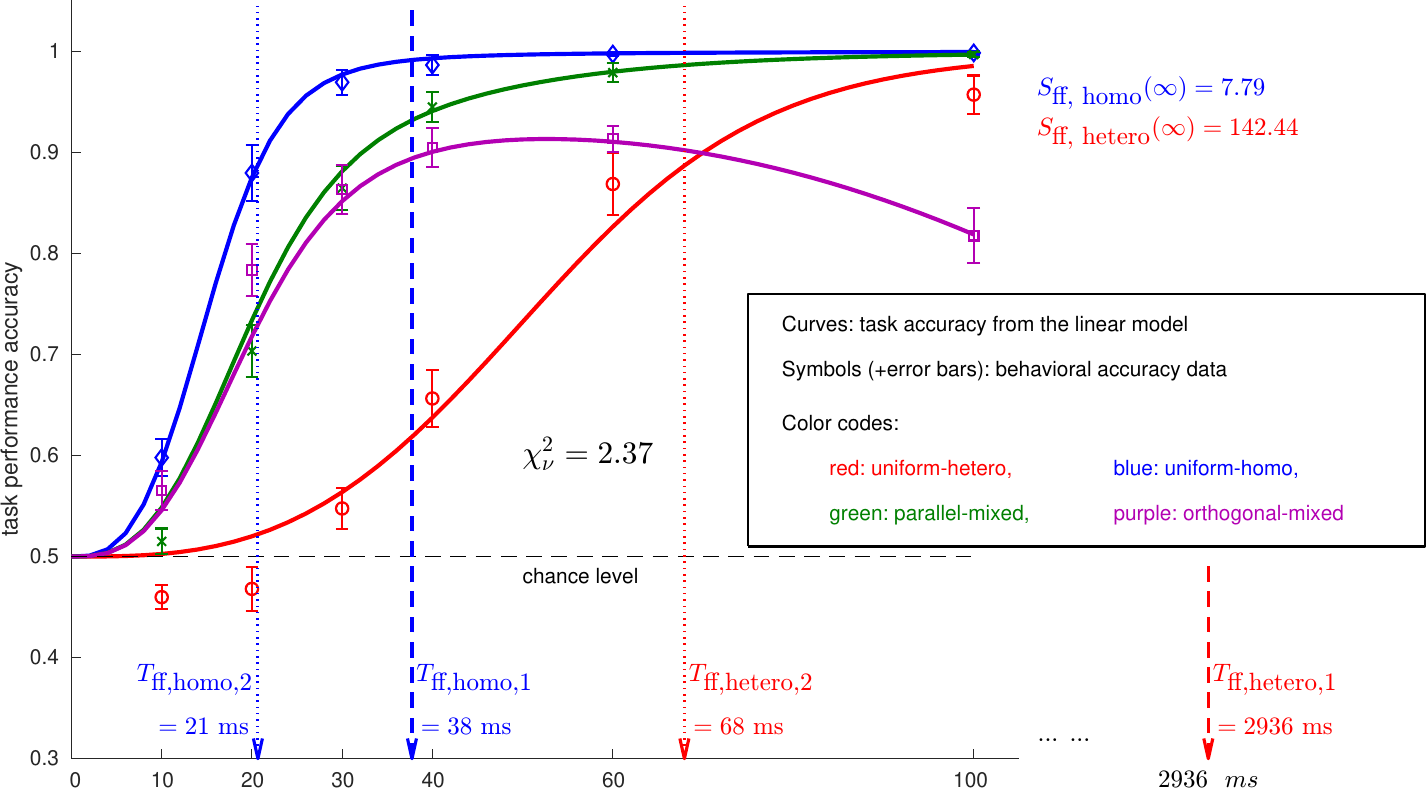}}
\put( 60, -5){\sf\scriptsize Stimulus duration $t$ (ms)}
\put(2, 210){\small \sf\textbf{Task performance accuracy by a super-linear model (curves) and by behavioral data }}
\end{picture}
\end{center}
\caption{\label{fig:feedforward_model}
Perceptual performance is poorly modeled by a super-feedforward model, which
employs only feedforward decision signals $S_{\textrm{ff}, i} (t) = S_{\textrm{ff}, i} (\infty ) (\textrm{V1}) \cdot 
 \left [1- \exp \left (- t/ T_{\textrm{ff}, i,1} \right ) \right ] \cdot 
 \left [1- \exp \left (- (t/ T_{\textrm{ff}, i,2})^2 \right ) \right ]$ 
using six parameters, 
three 
$(S_{\textrm{ff}, i} (\infty ), T_{\textrm{ff},i,1}, T_{\textrm{ff},i,2})$ for  each
$i=$ homo or hetero. The resulting $\chi^2_\nu = 2.373$ is about four times of the $\chi^2_\nu$ by the recurrent
model in Figure \ref{fig:model}A using the same number of parameters.  
This super-feedforward model cannot explain the flip tilt illusion, which led to accuracies that are lower than chance  for small $t$ 
on the uniform-hetero stimuli, and cannot explain the crossing of the accuracy curves for parallel-mixed and orthogonal-mixed stimuli.
The optimized temporal parameter values are marked on the figure.
}
\end{figure}

Our feedforward-feedback model uses six model parameters, three parameters each for its 
feedforward 
($S_{\textrm{ff},i} (t )$) 
and feedback 
($S_{\textrm{fb},i} (t )$) 
components in equations \ref{eq:FF} and \ref{eq:FB}.
Since a sufficient number of model parameters is necessary for the model to sufficiently agree with data,
we ask whether a super-feedforward model by six model parameters, but 
employing only a feedforward, one-stage, component, can explain our behavioral data equally well.
This super-feedforward model is made by modifying our feedforward-feedback 
model, in particular 
setting  
$S_{\textrm{fb},i} (t )=0$ 
and replacing
equation \ref{eq:FF} by 
\begin{align}
S_{\textrm{ff}, i} (t ) =& S_{\textrm{ff}, i}(\infty )  \cdot 
 \left [1- \exp \left (-  t/ T_{\textrm{ff},i,1}    \right ) \right ] \cdot 
 \left [1- \exp \left (- (t/ T_{\textrm{ff},i,2})^2 \right ) \right ], \nonumber \\
	& \mbox{for $i=$ homo or hetero}. \label{eq:SuperFF}
\end{align}
Hence, three parameters 
$(S_{\textrm{ff,homo}}(\infty)$, $T_{\textrm{ff,homo,1}}$, and $T_{\textrm{ff,homo,2}})$ 
are used to model $S_{\textrm{ff,homo}} (t )$,
and another three parameters analogously to model  
$S_{\textrm{ff,hetero}} (t )$,
giving six parameters in total.  
Regardless of whether $i$ is homo or hetero, 
information about V1's the whole population responses $(r_a, r_b, r_c, ...)$ (see Figure \ref{fig:fliptiltillusion_derive}) 
is transmitted to downstreams areas by a single sweep.
Processing this information downstream for the perceptual task takes time,
making decision signal $S_{\textrm{ff},i} (t )$ increase with time $t$.

When the set of six parameters for this super-feedforward model is optimized to 
minimize $\chi^2_\nu$, the result (shown in Figure \ref{fig:feedforward_model}) gives $\chi^2_\nu = 2.37$, 
which is about four times as large as the $\chi^2_\nu = 0.59$ by the feedforward-feedback model.
In particular, to explain the flip tilt illusion on the hetero-pairs, the model
needs to produce accuracies below the chance level. 
The super-feedforward model cannot do this,
nor can it account for the crossing between the accuracy curves for 
the parallel-mixed and orthogonal-mixed stimuli by the related reason.
(In addition, some of the optimized model parameters (marked in Figure \ref{fig:feedforward_model}) do not
appear to have suitably interpretations, suggesting the inadequacy of this
feedforward model. ) 
Critically, even when more model parameters are allowed, as long as there is no mechanism to make
$S_{\textrm{ff}, i}(t)$ negative, a below-chance accuracy in order to explain the flip tilt illusion is impossible.
Hence, it is essential to have two stages in the model to account for 
both below- and above-chance accuracies  for short and long $t$, respectively.
The first stage is the feedforward stage that has the partial and misleading information to make 
$S_{\textrm{ff, hetero}}(t)$ negative. The second stage is the feedback stage with additional 
queried information for more accurate visual perception.


\section{Summary and discussions}

In the looking-and-seeing framework for vision \citep{ZhaopingLookingAndSeeing2026}, 
the Central-Peripheral Dichotomy (CPD) theory \citep{ZhaopingNewFramework2019} 
states that
the peripheral visual field is mainly for looking \citep{ZhaopingPeripheral2024},  the central 
visual field is mainly for seeing, and the top-down feedback to aid seeing is mainly directed to the
central visual field.
This framework has predicted the flip tilt illusion, which has been confirmed to be typically 
only visible in the peripheral visual field \citep{ZhaopingFlipTilt2020}, because the peripheral 
field lacks the feedback to query for additional information to aid recognition in light 
of the bottleneck.  
This framework further predicts that this illusion should become visible 
in the central visual field when the top-down feedback is impaired.
The current study confirmed this prediction, using backward masking to disrupt the feedback.  
Furthermore, this study constructed a phenomenological model of the 
Feedforward-Feedback-Verify-reWeight (FFVW) algorithm of visual perception to account for our data.
This model  dissected the respective contributions by the feedforward and feedback processes in 
the orientation discrimination task to reveal the following.  Firstly, in this task, 
the feedforward processes alone can largely account for perception by a very short viewing 
duration $t < 20$ ms, making the flip tilt illusion visible in the 
central visual field.
Second, the feedback processes dominate the perceptual decision making for longer 
viewing durations $t> 20$ ms, vetoing the flip tilt illusion to make it invisible.

\subsection{Relating the flip tilt illusion with the reversed depth illusion}

\begin{figure}
\begin{center}
\setlength{\unitlength}{1.00pt}
\begin{picture}(500, 400)
\put(30, 0) {\includegraphics[width=0.87\textwidth]{ 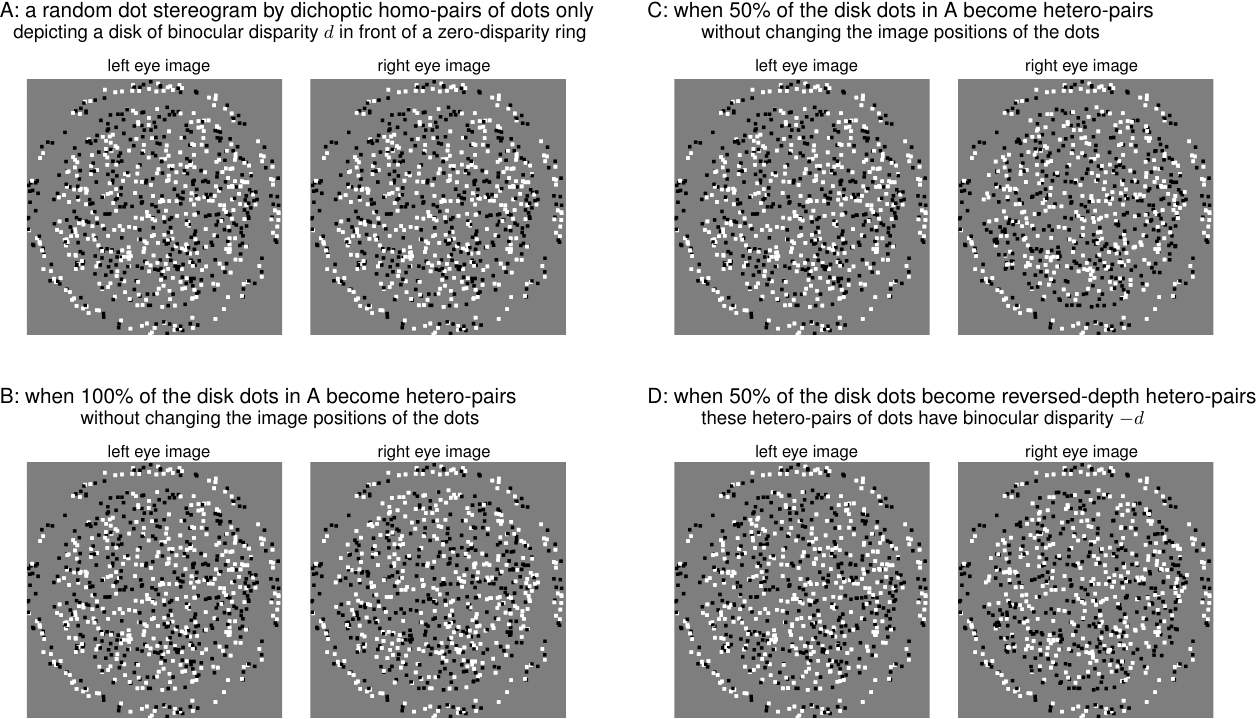}}
\end{picture}
\end{center}
\caption{\label{fig:Depth}
Reversed depth illusion is analogous to the flip tilt illusion and
offers additional insights.  
A, B, C, D are random dot stereograms using dichoptic homo- and hetero-pairs and
are analogous to uniform-homo, uniform-hetero, parallel-mixed, and orthogonal-mixed stimuli in Figure \ref{fig:Method}. A: a random dot stereogram depicting two depth surfaces in 3D space, one for a central disk, 
another for a surrounding ring.  Each surface is covered by randomly distributed black and white dots. 
Each dot in 3D space gives a dichoptic homo-pair of dots in the stereogram, one dot in each monocular image. 
Each dot on the ring has a zero disparity, and each dot on the disk has a positive disparity $d$ for a disk 
in front of the ring. Free fuse the two images to view the 3D scene.
B: a stereogram modified from A by reversing the contrast polarity, from black to white or vice versa, 
of each disk dot in the right eye image without changing any dot positions. 
The disk is now depicted by hetero-pairs of dots of disparity $d$.
Typical observers cannot judge the relative depth order between the disk and the ring.
C and D are made from A by randomly selecting 50\% of the disk dots and making them
hetero-pairs of dots, while the other 50\% of the disk dots remain unchanged (as homo-pairs).
In C, the contrast polarity of the selected disk dots was reversed in the right image only, 
without changing the dot positions, such that both the homo- and hetero-pairs of disk dots have 
disparity $d$.  D is made by modifying C as follows: for each hetero-pair of disk dots, swap
the monocular image positions of the two dots, reversing this dot's disparity to $-d$.
}
\end{figure}

This experimental confirmation of the theoretical prediction follows another
such confirmation of an analogous prediction by the same theory on another
illusion, the reversed depth illusion 
in stimuli when homo- and hetero-pairs are dichoptic  \citep{Zhaoping_iScience2025}. 
A dichoptic homo-pair consists of one dot each shown to the left and the right eyes, it
can arise from retinal images of a dot in the three-dimensional (3D) visual scene.
The  binocular disparity between the monocular image positions of the two
corresponding dots encodes the depth of the 3D dot relative to the zero disparity for 
the depth of the 3D fixation location.  
Figure \ref{fig:Depth}A is a random dot stereogram depicting a 3D scene 
containing a disk with disparity $d>0$ in front of a zero-disparity ring.
The surfaces of the disk and the ring are covered by randomly distributed black and white dots, 
each giving a dichoptic homo-pair of dots in the stereogram. 
These dots excite V1 neurons tuned to binocular disparities, 
leading to the perception of the disk in front of the ring.
When the monocular images are modified to make the disk homo-pairs into hetero-pairs 
without changing any dot positions (see Figure \ref{fig:Depth}B), 
V1 neurons respond as if the disparity $d$ becomes $-d$ \citep{CummingParker1997}, i.e., as if the disk is behind the ring.
This is analogous to how non-dichoptic hetero-pairs of dots
excite V1 neurons that prefer an orientation $90^o$ away from the veridical orientation (Figure \ref{fig:fliptiltillusion_derive}). 
As predicted by the CPD theory, observers can normally see this illusory reversed depth 
in the peripheral but not the central visual field \citep{ZhaopingAckermann2018}, just as the flip tilt illusion
is normally only visible in the peripheral visual field.
Furthermore, as predicted, this illusion becomes visible in the
central visual field by impairing the feedback query through backward masking \citep{Zhaoping_iScience2025}.

Meanwhile, the reversed depth illusion and the flip tilt illusion give 
different perspectives.
Whereas a non-dichoptic hetero-pair of dots has a visible and veridical orientation,
a dichoptic hetero-pair has no sensical depth, such that the feedback query in the
FFVW algorithm should only veto the feedforward depth signals without replacing
the feedforward signals by the alternative depth signals according to the disparity between 
the dichoptically corresponding dots.
Hence, when the viewing duration is sufficient for the veto, 
observers cannot discriminate whether the disk  
in Figure \ref{fig:Depth}B is in front or behind the ring.
This implies that  the asymptotic feedback decision signals,
 $S_{\textrm{fb, hetero}} (\infty )$  and $S_{\textrm{fb, homo}} (\infty )$,
depend not only on the binocular disparity but also on whether the two corresponding 
monocular dots are matched (homo-pair) or mismatched (hetero-pair) in their 
image contrast polarity \citep{DoiEtAl2011}.
This is in contrast to the orientation perception of  non-dichoptic dot pairs, 
when the same (or similar) value of 
$S_{\textrm{fb}}(\infty )=S_{\textrm{fb, homo}}(\infty ) = S_{\textrm{fb,hetero}}(\infty )>0$ 
applies  regardless of whether the dot-pairs are homo- or hetero-pairs.

In Figure \ref{fig:Depth}C and Figure \ref{fig:Depth}D, half of the disk dots 
are hetero-pairs while the other half remain homo-pairs, analogous to the parallel-mixed and orthogonal-mixed stimuli in 
Figure \ref{fig:fliptiltillusion_derive}.
In Figure \ref{fig:Depth}C,  
the homo- and hetero-pairs have the same disparity $d$.
In Figure \ref{fig:Depth}D,  
the homo-pairs have disparity $d$ and hetero-pairs have disparity $-d$.
Accordingly, in Figure \ref{fig:Depth}C or Figure \ref{fig:Depth}D, respectively, 
the preferred depth of neurons responding to the hetero-pairs 
disagree or agree with that of neurons responding to the homo-pairs for the stimulus 
i.e., the V1 depth signals evoked by the hetero-pairs are incongruent or congruent
with those evoked by the homo-pairs. 
Hence, the depth percept (of the disk in front of the ring) is clearer in 
Figure \ref{fig:Depth}D than in Figure \ref{fig:Depth}C, analogous to a higher
accuracy for the orthogonal-mixed than parallel-mixed stimuli
when $t < 30$ ms to make the feedforward mechanisms dominant
(see Figures \ref{fig:ResultSummary1} and \ref{fig:model}).

\subsection{The nonlinear constructiveness in the feedback mechanisms}

When viewing duration on the random dot stereograms is sufficient, if the feedback 
mechanisms completely veto the reversed depth signals evoked by the hetero-pairs, 
Figure \ref{fig:Depth}C and Figure \ref{fig:Depth}D should evoke the
same depth perception based on the same set of homo-pairs only.
However, the disk's depth remain better perceived in Figure \ref{fig:Depth}D than
in Figure \ref{fig:Depth}C, manifesting a non-linear constructive mechanism in perception.

With long duration viewing in the central visual field, 
typically observers cannot at all discriminate the disk's depth when all 
the disk's dots are hetero-pairs in Figure \ref{fig:Depth}B.
This implies that, combining the feedforward and feedback contributions as $t\rightarrow \infty$,
the decision signal on the hetero-pairs is nearly zero, i.e.,  $S_{hetero}(t\rightarrow \infty) \approx 0$.
However, when hetero-pairs are mixed with homo-pairs for the disk 
in Figure \ref{fig:Depth}CD,  the disk's depth (according to the disparity of the homo-pairs) appears 
clearer when they are congruent (in  Figure \ref{fig:Depth}D) rather than 
incongruent ( Figure \ref{fig:Depth}C) with the homo-pairs in their evoked V1 signals. 
This implies that, although the illusory depth signals from V1 about 
the hetero-pairs are often vetoed, they are constructively used 
for perception when they are congruent with the verifiable depth signals from the 
homo-pairs.  Indeed, a previous study \citep{ZhaopingHeteroHomoDots_RDS2021} has shown
the following: with a sufficiently long viewing duration, incongruent hetero-pairs (when mixed with homo-pairs)
are perceptually equivalent to dichoptically non-corresponding noise, whereas the 
congruent hetero-pairs augment the depth signals from the homo-pairs.

In the current study with non-dichoptic dot-pairs, an analogous nonlinear constructive use of the illusory 
feedforward signals from the hetero-pairs is also noticeable.
In particular, at $t = 20$ ms, the performance for
the uniform-hetero condition was approximately at
the chance channel (see Figure \ref{fig:ResultSummary1}A), implying a zero decision signal 
$S_{hetero}(t) \approx 0$. 
By our linear model (equations \ref{eq:Sparallel-mixedDefinition} and
\ref{eq:Sorthogonal-mixedDefinition}), 
  $S_{\textrm{parallel-mixed}} = (S_{\textrm{homo}}+ S_{\textrm{hetero}})/2$
and 
  $S_{\textrm{orthogonal-mixed}} = (S_{\textrm{homo}}- S_{\textrm{hetero}})/2$,
one would expect $S_{\textrm{parallel-mixed}} \approx S_{\textrm{orthogonal-mixed}}$.
Nevertheless, at this $t$, the accuracy for the orthogonal-mixed 
(whose hetero- and homo-pairs give congruent V1 signals) was 
significantly ($p<0.001$) higher than that for the parallel-mixed stimuli
(see Figure \ref{fig:ResultSummary2}B). 
At $t  = 30$ ms, the accuracy for uniform-hetero was close to, but significantly higher than, 
the chance level, implying that $S_{hetero}(t) >0$. This would imply by our model 
that the accuracy for the parallel-mixed should be higher than that for orthogonal-mixed. 
Nevertheless, in our data, the accuracy for orthogonal-mixed was statistically equivalent to that for parallel-mixed.
In both sets ($t=20$ and $30$ ms) of data points, the effects of the illusory depth signals 
from V1 neural responses to the hetero-pairs are stronger than expected from our linear model, 
manifesting the constructive process that our parsimonious model does not yet capture,
 even though on average across the  24 data points our model performs
reasonably well (Figure \ref{fig:model}). 
The constructive process could be modelled by a nonlinear term such as 
$\alpha \cdot S_{\textrm{ff, homo}} \cdot S_{\textrm{ff, hetero}}$ to model 
the interaction between the homo- and hetero-pairs,
at the expense of an extra model parameter $\alpha$ and thus the parsimony of the model. 
This could be explored further in a future study.

Constructive use of fragmentary or even illusory feedforward signals is a very useful 
property of visual perception under the constraint of an information bottleneck, and deserves
deeper investigations in future studies.
The Kanizsa triangle in Figure \ref{fig:FeedforwardFeedback}C arises from 
the constructive use of the fragmentary image information alluding to the 
triangle, under a perceptual hypothesis about the presence of the white occluding triangle,
the black disks, and the occluded triangle with a black outline.
This perceptual hypothesis is then verified by the feedback query using analysis-by-synthesis. 
The fragmentary image signals about the occluding and occluded
objects interact with each other nonlinearly, such that the perception of one object
augment that of another object.
Neural responses to illusory contours indeed emerge in V1 after they appear 
in V2 \citep{LeeNguyen2001}.
Perception of such illusory occluding surfaces is indeed less vivid in 
the peripheral visual field which lacks the feedback \citep{MooreEtAl2025}.
Perception of vivid objects (e.g., faces) in Mooney or 
two-tone images likely involves similar constructiveness.

\subsection{Modeling the FFVW algorithm without specifying its neural implementations}

Our model of the FFVW algorithm is a phenomenological model, which emphasizes the
computational principle and the algorithmic design without specifying the neural 
implementation details.  Future works should find out the neural implementation, such
as to identify the source brain areas for the feedback signals and the postsynaptic 
target cell types of the feedback. Nevertheless, this model can already provide
fresh and quantitative information about the feedforward and feedback processes for visual perception.

In particular, each of the six model parameters have its 
computational and algorithmic counterparts.
\begin{enumerate}
\item   
$S_{\textrm{ff,homo}}(\infty )$ and 
$S_{\textrm{ff, hetero}}(\infty )$ 
capture the strengths of the most significant V1 neural responses to a 
homo- and hetero-pair of dots in the initial feedforward signals through 
the information bottleneck, and capture the identity or sign of the feature line 
labels (whether it is clockwise or counterclockwise tilted from vertical) 
in the feedforward signals for the orientation perception.

\item 
$T_{\textrm{ff}}$ captures the time scale for any 
initial feedforward signals to impact the perceptual hypotheses on the
orientation discrimination task, regardless of whether these signals are from 
homo-pairs or hetero-pairs.

\item 
$S_{\textrm{fb}}(\infty)$ captures the strength of an orientation label (whether
it is clockwise or counterclockwise tilted from vertical) of the axis of alignment
in a recognized pair of dots, regardless of whether the dot-pair is a homo- or 
a hetero-pair.

\item 
$T_{\textrm{fb,homo}}$ captures the time scale since stimulus onset 
for the feedback process to query for additional sensory information and to verify 
and confirm the leading perceptual hypothesis initiated by the initial feedforward signals.

\item   
$T_{\textrm{fb,hetero}}$ is computationally similar to, but should be numerically longer 
than, $T_{\textrm{fb,homo}}$, since it should take longer to veto the initially 
leading hypothesis (the flip tilt illusion) by the impoverished feedforward 
information and to establish an alternative hypothesis 
(using the additional information queried by the feedback) for the perceptual outcome.
\end{enumerate}

The six parameters are sufficient to capture some essence of the FFVW algorithm, so that fitting the model to our data has provided 
the following insights.  First, the parameter $T_{\textrm{ff}} =16.6$ ms suggests that
the feedforward processing is substantially completed within about $20$ ms.  
This value can be related to a $20$ ms difference between the response onset 
latencies of neurons in V1 and those in the frontal eye field 
(FEF) \citep{NowakBullier1997,SchmoleskyEtAl1998}.
Our data indicate that visual recognition by only $10$ ms can be better 
than chance. This agrees with a previous finding that observers can detect meaning
 in rapid serial visual presentations of pictures at $13$ ms per picture \citep{PotterEtAl2014}. 

Second, the parameter $T_{\textrm{fb,homo}} = 35.4$ ms
suggests that the feedback process can be very fast, lagging the feedforward
process by only about 20 ms.  This is comparable with a feedback lag of
$10$ to $40$ ms from MT to V1 suggested by a transcranial magnetic 
stimulation (TMS) study \citep{Pascual-LeoneWalsh2001}, 
$30$ to $40$ ms from V4 to V1 from neurophysiological studies\citep{ChenYanEtAl2014, KlinkEtAl2017, XinEtAl2025}, 
$11$ ms from magnetoencephalographic recordings \citep{BoehlerEtAl2008}, and
$\sim 20$ ms inferred from activity waves in the cortex measured by voltage 
sensitive dye \citep{RolandEtAl2006}.  
Since $T_{\textrm{fb,homo}}$ is comparable to $2\cdot T_{\textrm{ff}}$, 
it suggests that one iteration of the feedback query is sufficient when the 
perceptual outcome is consistent with the initial leading hypothesis
suggested by the feedforward signals admitted into the bottleneck. 

Third, since the  parameter $T_{\textrm{fb,homo}}  = 90.2$ ms for the hetero-pairs 
is much longer than  $T_{\textrm{fb,homo}}$ for the homo-pairs, 
the feedback process should be  complex, depending on the stimulus complexity, 
computations involved (e.g., whether to confirm or veto the initially leading 
perceptual hypothesis), and the task. 
Indeed, the range of feedback lags reported in the literature is 
large \citep{WyatteEtAl2014} and, when objects are more difficult to recognize 
due to factors such as occlusion, it takes longer for object information to 
emerge in responses of neurons in inferotemporal cortex \citep{KarEtAl2019}.  
Since  $T_{\textrm{fb,homo}}$ is multiple times $ T_{\textrm{ff}}$, it is likely
that many iterations of the feedback query are needed when perception is 
challenging, such as when vetoing and revising the initial perceptual hypotheses 
are needed for sufficiently confident perceptual outcome.
When the viewing duration $t$ is self-regulated rather than externally limited,
how the cognitive system metacognitively determines this duration, or the number of feedback queries,  
to make a perceptual decision
is an active research area \citep{BotvinickEtAl2004, ChurchlandEtAl2008, Mamassian2016, SchulzEtAl2023}.

Fourth, comparing $T_{\textrm{fb,homo}} = 35.4$ ms with  
$T_{\textrm{fb,homo}}  = 90.2$ ms suggests that easy visual discrimination 
tasks will not be impaired by backward masking if the mask appears $\ge 40$ ms 
after stimulus onset, whereas more difficult or non-trivial discrimination 
should be substantially impaired by such masking unless the mask appears 
after around $60$ to $100$ ms. This is consistent with 
experimental data\citep{CauchoixEtAl2016, BreitmeyerOgmen2006}. 
Our model suggests that whether a visual discrimination task can be
substantially impaired or not by a mask that appears $\sim 50 $ ms after
stimulus onset is determined by whether the ultimate perceptual outcome
without masking disagrees or agrees with the leading perceptual hypothesis
from the initial feedforward signals through the bottleneck.

Fifth, the magnitude $|S_{\textrm{fb}}(\infty )|$ for the feedback decision
signal is much larger than those, 
$|S_{\textrm{ff, homo}}(\infty)|$ and 
$|S_{\textrm{ff, hetero}}(\infty )|$, for the feedforward
decision signals. This, together with $T_{\textrm{fb,homo}} = 35.4$ ms, suggests 
that feedback mechanisms provide dominant contribution to perceptual outcomes 
unless the viewing duration is less than $20$ ms, even 
for visual stimulus for which discrimination is not severely
impaired by typical masking (with mask onset $\ge 40$ ms after stimulus onset).
As natural fixation durations are $\sim 200$ ms, our findings suggest
that feedback mechanisms dominate perception of foveated objects 
during fixation.  

Traditionally,  visual inputs within the first $\sim 150$ ms have been viewed as processed mostly
by feedforward mechanisms \citep{DiCarloEtAl2012, CauchoixEtAl2016}. This view has
inspired artificial neural networks for visual recognition using mainly 
feedforward neural networks, and has inspired the idea that primate visual system 
could be largely understood through such artificial neural networks \citep{SerreWolfEtAl2007, YaminsEtAl2014}.
Our findings question these traditional views.

\subsection{Contrasting the central versus the peripheral visual field in their perceptual processes} 

Our experimental stimuli, uniform-homo, uniform-hetero, parallel-mixed, and orthogonal-mixed (see Figure \ref{fig:Method}), correspond to
those shown in A, D, C, and B in our Figure \ref{fig:fliptiltillusion_demo} to demonstrate the flip tilt illusion.
Our experimental stimuli were viewed in the central visual field, and 
when viewing duration $t$ is sufficiently long to suppress the flip tilt illusion by the feedback mechanisms,
performance for the parallel-mixed stimulus was higher than that for orthogonal-mixed.
In contrast,  the ring in Figure \ref{fig:fliptiltillusion_demo}B (which corresponds to our orthogonal-mixed stimlus) appears more apparent than 
Figure \ref{fig:fliptiltillusion_demo}C (which corresponds to our parallel-mixed stimlus) when we casually view them for a duration much longer than
$100$ ms, demonstrating that the flip tilt illusion is effective even for $t\gg 100$ ms.  
This is because the rings in Figure \ref{fig:fliptiltillusion_demo} are viewed mostly in our peripheral visual field.  Even when we fixate our gaze to a segment 
of a ring, the other segments of the ring are in the peripheral visual field. According to the CPD theory, the top-down feedback query is absent or much 
weaker in the peripheral visual field.  Hence, no restriction on duration $t$ is necessary 
to see the flip tilt illusion in such a peripheral ring stimulus. 
By the CPD theory \citep{ZhaopingNewFramework2019, ZhaopingLookingAndSeeing2026}, vision in the peripheral field is mainly for 
looking rather than seeing \citep{ZhaopingPeripheral2024},  while vision in the central visual field is mainly for seeing. 
Hence,  the central and peripheral fields work as a team for vision.

\subsection{Relationship between visual crowding and the illusions in the peripheral visual field}

Like the flip tilt illusion and reversed depth illusion, visual crowding is also
typically present in the peripheral but not the central visual field, as is demonstrated in Figure \ref{fig:FeedforwardFeedback}C.
Traditionally,  crowding has been studied as a visual perceptual phenomenon, with extensive efforts to 
understand its underlying neural or phenomenological mechanisms 
\citep{Pelli2008, LeviCrowdingReview2008, BalasEtAl2009, WhitneyLevi2011, ChenEtAl2014, MillinEtAl2014, ManassiWhitney2018, 
RosenholtzEtAl2019, SchwetlickHerzog2025}.
According to the CPD theory, crowding occurs due to the following two reasons. 
First, a bottleneck impoverishes the information available for for inferring or decoding visual inputs sufficient well. 
Second, the peripheral visual field has no feedback mechanisms to query for additional information in light 
of the bottleneck  \citep{ZhaopingNewFramework2019, ZhaopingPeripheral2024}.
The first reason has been duly noted in previous studies on crowding. 
The CPD theory offers the second reason in addition.  
By this second reason, perceptual ambiguity due to crowding in the peripheral visual field 
cannot be overcome simply for viewing the stimulus for a longer duration to accumulate 
additional information by feedback query.  

Hence, although crowding is apparently very different from the illusions, it offers another 
window to understand visual perceptual processes through the CPD theory, which provides falsifiable predictions.
One such prediction is that crowding should also occur in the central visual field if viewing duration $t$ 
(terminated by a mask) is sufficiently short to make feedback query ineffective \citep{ZhaopingV1Function2026}.  
Our model of FFVW on the flip tilt illusion, although not on crowding, nevertheless suggests 
that, when a visual stimulus is easier to recognize, then the viewing duration $t$ should be shorter to 
induce crowding in the central visual field.
A related prediction is a weaked or absent feedback to the representation of the central visual field 
in early visual stages such as V1 for the amblyopic eyes, since foveal crowding is present in perception by amblyopic eyes \citep{Levi2020}.
Similarly, a less developed mechanisms for the feedback query is predicted to be a
reason for stronger crowding, particularly near or in the central visual field, in human children than adults \citep{WaughEtAl2026}.

\subsection{Relationship with previous works on feedbacks along the visual pathway for vision}

It has been decades and centuries since the idea emerged that perception uses analysis-by-synthesis computation and top-down 
feedback \citep{Al-Haytham1024, Helmholtz1925,  Mackay1956, HalleStevens1959, 
CarpenterGrossberg1987ART, LiOdorAdaptation1990, KawatoEtAl1993, DayanEtAl1995, HochsteinAhissar2002, 
YuilleKersten2006, BeverPoeppel2010, GilbertLi2013, RoelfsemaEtAl2016}, applying to visual, auditory, olfactory, 
and even multisensory perception \citep{ZhaopingCPD2023}.
The new elements in the CPD theory of the looking-and-seeing framework include
the following. First, the feedback is mainly aimed at the representation of the central 
visual field. Second, the feedback is motivated by a bottleneck and aims to query
for additional information in order to aid recognition in light of the bottleneck.
Although synthesizing the would-be inputs  according to the candidate hypotheses about the scene
is part of the mechanisms for the feedback to query for information, 
the synthesis is aimed at analyzing the sensory inputs (through hypotheses testing, hence analysis-by-synthesis)
rather than predicting the sensory inputs (such as in predictive encoding).
Third, the CPD applies multisensorily, such that, e.g., primate auditory inputs can function peripherally to select
sensory inputs to guide gaze towards the auditory source, and visual inputs can guide rodents
to orient its ``foveal" nose and microvibrassae \citep{ZhaopingCPD2023}.

Some previous works have identified or inferred feedback from representations of the peripheral visual field 
in higher visual areas to that of the central visual field in early visual stages such as V1 
along the visual pathway \citep{WilliamsEtAl2008, FanEtAl2016, KnapenEtAl2016, KroellRolfs2025, HouHeZhang2026}. 
They are agnostic regarding whether top-down feedback to the central visual field is denser
than the feedback to the peripheral visual field.  Meanwhile, in the CPD theory, 
the central field focus of the feedback applies also to trans-saccadic situations to link a peripheral location 
and a central location by a saccade \citep{ZhaopingNewFramework2019, ZhaopingLookingAndSeeing2026}. 
In such a situation, a target in a peripheral location is the target of attention and gaze shift, 
such that its post-saccadic location is in the central visual field. To recognize this target's feature,  
the initial feedforward information arises from the target's initial location in the peripheral field, 
and the subsequent feedback query is aimed at the target's expected, upcoming, post-saccadic, location in the central visual 
field.i  Such a trans-saccadic feedback manifests a phenomenologically a feedback query from the 
representation of a peripheral visual location to that of a central location, and is similar or
related to those in the previous works, is also related to works on transsaccadic integration, perceptual stability, 
and object based attention \citep{Irwin1991,StewartSchuetz2018, Wurtz2018,WangEtAlQian2024,LiangZhaoping2025,CavanaghMelcher_CurrentOpinion2026}.

\subsection{Concluding remark}

	The flip tilt illusion is an example of visual perceptual phenomena that enable multiple
perspectives on the visual perceptual process through a bottleneck \citep{ZhaopingLookingAndSeeing2026}.
First, this illusion is typically  visible only in the peripheral visual field, as predicted by
the CPD theory \citep{ZhaopingFlipTilt2020}, because perception is misled by impoverished and misleading information 
about visual inputs fed forward through the bottleneck and because the peripheral visual field does not 
have the feedback mechanism to query for additional visual input information to correct the illusion.
Second, in the central visual field which has the privilege of the feedback query, this illusion
becomes visible when the feedback query  is made ineffective by backward masking, as predicted previously
and confirmed by the study reported here.
Third, through various types of stimuli made by combining homo-pairs and hetero-pairs
in an orientation discrimination task, and through an appropriately formulated computational model, 
the flip tilt illusion provides a glimpse into the temporal dynamics and relative contributions of 
the feedforward and feedback mechanisms in the FFVW algorithm.

Future work should try to apply this approach, particularly the computational modeling, 
to other visual stimuli in other perceptual tasks. 
The model parameter values and precise details of the model formulation will have to depend on the visual stimuli and 
on the task.  
For example, for depth discrimination task using dichoptic dot-pairs, the feedback decision signal $S_{fb}$ is unlike that for the non-dichoptic dot-pairs in an 
orientation discrimination task, such that this decision signal depends on whether the dot-pairs are homo- or hetero-pairs.
These model formulation details and parameter values will nevertheless provide constraints or hints on the underlying 
neural implementations for the FFVW algorithm.  Future studies should design various other example stimuli and example tasks 
that are informative about various aspects of neural mechanisms for visual perception.

\section*{Acknowledgements}
This work was supported by funding from the University of T\"ubingen and the Max Planck Society.  

Declaration of commercial interests and conflicts: none

\bibliographystyle{elsarticle-harv} 
\bibliography{ReferenceForBook.bib}

\end{document}